\pdfoutput=1

\documentclass{article}[11pt,a4paper]
         
\usepackage{graphicx}
\usepackage[english]{babel}
\usepackage[fleqn]{amsmath}
\usepackage{amssymb}
\usepackage{bm}
\usepackage{tikz}
\usepackage{lmodern,microtype}
\usepackage[left=4cm]{geometry}
\usepackage{hyperref}

\newcounter{conjecture}
\newenvironment{conjecture}[1][]{\refstepcounter{conjecture}\par\medskip\noindent%
   \textbf{Conjecture~\theconjecture. #1} \rmfamily}{\\ \medskip}

\renewcommand{\i}{\text{i}}

\newcommand{\tr}{\mathop{\text{tr}}}
\newcommand{\sgn}{\mathop{\text{sgn}}}

\title{The eight-vertex model and lattice supersymmetry}
\date{}
\author{Christian Hagendorf\footnote{Section des Math\'ematiques,
Universit\'e de Gen\`eve, 2-4 rue du Li\`evre, CP 64, 1211 Gen\`eve 4,
Switzerland, and Kavli Institute for Theoretical Physics, University of California, Santa Barbara, CA 93106. E-mail:
\href{mailto:christian.hagendorf@unige.ch}{christian.hagendorf@unige.ch}},
and Paul Fendley\footnote{University of Virginia, Department of
Physics, 382 McCormick Rd, Charlottesville, VA 22904, and Microsoft Station Q, University of California, Santa Barbara, CA 93106. E-mail:
\href{mailto:fendley@virginia.edu}{fendley@virginia.edu}}}

\begin{document}
\maketitle

\begin{abstract}
We show that the XYZ spin chain along the special line of couplings
$J_xJ_y+J_xJ_z+J_yJ_z=0$ possesses a hidden $\mathcal N=(2,2)$
supersymmetry. This lattice supersymmetry is non-local and changes the
number of sites. It extends to the full transfer matrix of the
corresponding eight-vertex model. In particular, it is shown how to
derive the supercharges from Baxter's Bethe ansatz. This analysis
leads to new conjectures concerning the ground state for chains of odd
length. We also discuss a correspondence between the spectrum of this
XYZ chain and that of a manifestly supersymmetric staggered fermion
chain.

\end{abstract}
\newpage

\tableofcontents

\section{Introduction}

The solution of the zero-field eight-vertex model by Baxter is a
landmark in the theory of exactly solvable systems. The seminal papers
\cite{baxter:72,baxter:72_2,baxter:73,baxter:73_2,baxter:73_3} present
a variety of algebraic and analytic tools to compute the partition
function, the eigenvalues and eigenvectors of its transfer matrix. The
model is still under active study: in particular, Fabricius and McCoy
showed that at the so-called root-of-unity points the spectrum of the
transfer matrix possesses degeneracies not easily explained by its
standard integrability alone
\cite{deguchi:02,fabricius:03,fabricius:05,fabricius:06}. It is
natural to attribute them to the presence of extended symmetries,
possibly elliptic generalisations of the $\text{sl}_2$-loop-algebra
symmetries discovered for the six-vertex model
\cite{fabricius:01,deguchi:01,korff:04}.

The topic of this paper is the eight-vertex model at a particular
root-of-unity point with an extended symmetry. We utilise the quantum
XYZ spin chain, whose Hamiltonian commutes with the
eight-vertex model transfer matrix \cite{sutherland:70}. More precisely, we
study the Hamiltonian
\begin{subequations}
\begin{equation}
  H_{N}= -\frac{1}{2}\sum_{j=1}^N \left(J_x\sigma_j^x\sigma_{j+1}^x+J_y\sigma_j^y\sigma_{j+1}^y+J_z\sigma_j^z\sigma_{j+1}^z\right) + \frac{N (J_x+J_y+J_z)}{2}
\end{equation}
with periodic boundary conditions along the special line of couplings
\begin{equation}
  J_xJ_y+J_xJ_z+J_yJ_z=0.
\end{equation}
 \label{eqn:xyzham}
\end{subequations}

While it was already noticed in Baxter's original works that along the
line of couplings \eqref{eqn:xyzham} the ground state energy per site
remains zero in the thermodynamic limit, its finite-size ground state
was addressed much more recently. In
\cite{stroganov:01,stroganov:01_2} Stroganov argued that
\eqref{eqn:xyzham} possesses exactly two zero-energy ground states for
$N$ odd.  When $J_x=J_y$ (so that $\Delta\equiv J_z/J_x= -1/2$), the
resulting critical XXZ chain has been extensively studied over the
past decade \cite{razumov:00, razumov:01,degier:02,fendley:03,
yang:04, razumov:04, difrancesco:04,difrancesco:04_2}.  The components
of the ground state in the basis with $S_z$ diagonal possess some
remarkable properties. They display a variety of relations with
combinatorial quantities, such as the enumeration of alternating sign
matrices and plane partitions. Many of the early conjectures, such as
sum rules for the components, were then proved with the help of
techniques such as the qKZ equations, and combinatorial tools
\cite{difrancesco:05_3,difrancesco:06, razumov:07,cantini:10}.  

The fact that the ground state energy is exactly zero in the XXZ chain
at $\Delta=-1/2$ for a finite odd number of sites was proved by
exploiting a hidden supersymmetry \cite{fendley:03,yang:04}. A
ground-state energy of exactly zero is a common characteristic of
theories with supersymmetry \cite{witten:82}.  While quite a number
of lattice models possess scaling limits described by field theories
with supersymmetry \cite{saleur:93_2}, only a few are known where the
supersymmetry is explicitly present on the lattice. Here, an
unusual feature is that the supersymmetry operator changes the number
of sites by one. While unusual, it is not unheard of; 
similar operators were for example studied in a spin chain arising from the
integrable structures in four-dimensional gauge theory \cite{beisert:08}.

Remarkable properties of the zero-energy ground state persist along
the entire line (\ref{eqn:xyzham})
\cite{bazhanov:05,bazhanov:06,mangazeev:10,razumov:10,fendley:10}. Using
the convenient parametrisation
\begin{equation}
  J_x=1+\zeta,\,J_y=1-\zeta, \, J_z = (\zeta^2-1)/2,
  \label{eqn:param}
\end{equation}
computer results indicate that the ground state at odd $N$ can be
expressed as polynomials in $\zeta$ with positive integer
coefficients. Non-linear recursion relations were observed for several
components of the zero-energy ground states of \eqref{eqn:xyzham}
\cite{bazhanov:05,bazhanov:06,mangazeev:10} (see also
\cite{rosengren:09}). These relations are described by the
tau-function hierarchies of the Painlev\'e VI equation
\cite{okamoto:87}.  Moreover, very simple expressions for one-point
functions in the ground state such as the magnetisation were found by
summing the expansion around the trivially solvable point
$\zeta\to\infty$ \cite{fendley:10}.

Since the ground-state energy remains zero along the entire line
(\ref{eqn:xyzham}), it is natural to expect that the supersymmetry
persists off the critical point. This is true in the scaling limit,
where the XYZ chain is described by the sine-Gordon field theory. This
field theory possesses symmetries that can be related to the affine
quantum group $U_q(\widehat{\text{sl}}(2))$ with zero centre
\cite{bernard:90}. When $q^2=-1$, the quantum group contains the
$\mathcal N=(2,2)$ supersymmetry algebra. This is precisely the value
of $q$ corresponding to the scaling limit along the line
(\ref{eqn:xyzham}). We thus refer to this line henceforth as the
``supersymmetric'' line.

The purpose of this paper is to show that the supersymmetry remains
exact on the lattice all along the supersymmetric line.  The main
focus in most (but not all \cite{deGier:05}) earlier studies was the
ground state of the system, which has exactly zero energy for any odd $N$.
In this article, we study not only the ground states of this XYZ spin
chain, but also consider its full spectrum.  We show that for any
value of $\zeta$, in certain momentum sectors this model possesses a
hidden exact symmetry which we call lattice supersymmetry, a lattice
version of the well-known $\mathcal N=2$ supersymmetry algebra
\cite{witten:82}. These results provide a systematic generalisation of
the XXZ results to the XYZ setting.

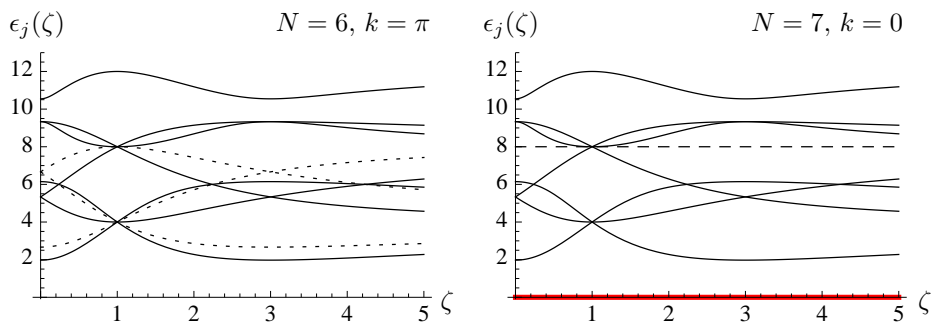
\begin{figure}[h]
  \centering
  \begin{tikzpicture}
    \draw (0,0) node {\includegraphics[width=0.9\textwidth]{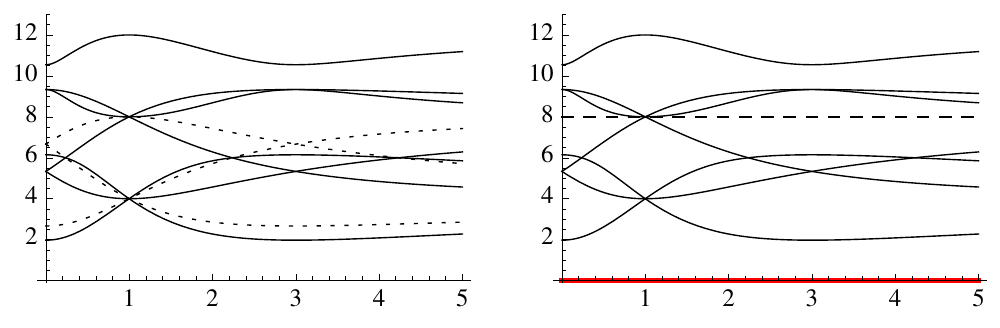}};
    \draw (-0.15,-1.5) node {$\zeta$};
    \draw (6.2,-1.5) node {$\zeta$};
    \draw (-5.6,2.2) node {$\epsilon_j(\zeta)$};
    \draw[xshift=4.2cm] (-5.6,2.2) node {$N=6,\, k=\pi$};
    \draw (0.7,2.2) node {$\epsilon_j(\zeta)$};    
    \draw[xshift=4.2cm] (0.7,2.2) node {$N=7,\, k=0$};    
  \end{tikzpicture}
  \caption{Rescaled eigenvalues $\epsilon_j(\zeta)=
  4E_j(\zeta)/(3+\zeta^2)$ for the XYZ Hamiltonian with $N=6$ and
  $N=7$ sites along the supersymmetric line as a function of
  $\zeta$. The plots are restricted to the subsectors with momentum
  $k=\pi$ and $0$ respectively. The solid lines correspond to exact
  common eigenvalues for the two chains. Moreover, the plot for $N=7$
  shows the existence of exact zero energy ground states.}
  \label{fig:spectra}
\end{figure}

We illustrate this symmetry by explicitly diagonalising the XYZ
Hamiltonian along the supersymmetric line. Figure \ref{fig:spectra}
shows the example of chains with $N=6$ and $N=7$ sites: in particular
momentum sectors, the corresponding Hamiltonians sites have exact
common non-zero eigenvalues. This is a well-known feature of supersymmetric
theories: states with $E>0$ are doubly degenerate. Thus, in our case
the common eigenvalues are candidates for supersymmetry
doublets. This is a consequence of the existence of supercharges which
change the number of sites $N$ by one, and serve as intertwiners for
the Hamiltonians of the corresponding chains. 
Moreover, two exact zero-energy ground states are found for odd $N$,
and they are potential candidates for so-called supersymmetry singlets (who do not have a superpartner). In fact,
we will show that in the sectors with momentum zero for odd $N$ and
momentum $\pi$ for even $N$, the states organise themselves into
quadruplets: an eigenvalue occurring first in the spectrum for $N-1$
sites, appears twice at $N$ sites, and once at $N+1$ sites. The
quadruplet structure hints at the existence of two copies of the
supersymmetry algebra. Indeed by taking into account the symmetry
under flipping all spins, we find a ${\cal N}=(2,2)$ supersymmetry
algebra on the lattice.

The transfer matrix of the eight-vertex model commutes with the XYZ
Hamiltonian, so the two have the same eigenstates. Thus one might
expect that these properties carry over to the full model. As we will
show, the existence of the $\mathcal N=(2,2)$ supersymmetry algebra is
deeply rooted in the eight-vertex model, and can be derived from the
Bethe ansatz. Moreover, the Bethe ansatz analysis leads to
a novel characterisation of the ground states of the XYZ chain \eqref{eqn:xyzham}
for odd $N$.

The XYZ chain is closely connected to another model with lattice
supersymmetry, introduced in \cite{fendley:03_2} and thoroughly
analysed in \cite{fendley:03,huijse:10,huijse:11_2}.  It
describes spinless fermions on a chain with nearest-neighbour as well
as the usual on-site exclusion. Connections between the two at the
critical point are already known; energy levels coincide with those of
the XXZ chain at $\Delta=-1/2$ with a momentum-dependent twist
\cite{fendley:03}. A similar correspondence also holds in the open
chain with a magnetic field \cite{yang:04}. Moreover, the ground states
of the fermion models display a variety of relations with the combinatorics
describes above \cite{beccaria:05}. Therefore, it is natural to ask if a
similar equivalence holds off the critical point. Indeed, several
connections between the zero-energy ground states for chains of odd
length and the ground states of a staggered version of the fermion
models were observed in \cite{fendley:10,fendley:10_1}. In this work,
we provide more evidence for the connection of the two models. We
observe the systematic existence of common eigenvalues in their
spectra after proper identification of the relevant parameters, even
in sectors where the supersymmetry is not fully realised.

The layout of this paper is the following. In section
\ref{sec:hamiltonian}, we introduce supercharges for the XYZ chain
which change the number of sites and study their properties. In
particular, we show that Hamiltonian can be obtained as a quadratic
form of the supercharges in special momentum sectors. The combination
with various other symmetries of the XYZ chain will lead to a second
lattice supersymmetry, and thus explain the quadruplet structure in
the spectra. In section \ref{sec:bethe}, we change our point of view
and study the system using Bethe ansatz for the eight-vertex
model. After recalling Baxter's original approach, we present a
derivation of the supersymmetry from the Bethe equations, and show
that it is a symmetry of the full transfer matrix along the
supersymmetric line, connecting systems of different sizes. After this we
proceed with a conjecture on the nature of the ground states for
chains of odd length. The relation of
the present model to the theory of fermions with nearest-neighbour
exclusion is discussed in section \ref{sec:fermions}: we observe that
upon proper identification of the relevant parameters, the spectra of
both theories have exact common eigenvalues. We suggest a mapping
between the two models, based on the Bethe ansatz solution of the
eight-vertex model. Finally, we present our conclusions and various
open problems in section \ref{sec:concl}. Some technical details are
relegated to appendices.

Throughout the paper we report observations obtained from exact
diagonalisation of small system sizes. Most of them show patterns
which seem to hold for general $N$, and thus we formulate them as
conjectures.

\section{Supercharges and the XYZ Hamiltonian}
\label{sec:hamiltonian}
In this section, we establish the relation between the XYZ chain
\eqref{eqn:xyzham} and lattice supersymmetry. We start with a review
of elementary symmetries of the XYZ chain and then proceed to the
definition of the supercharges. The analysis of their properties under
translation allows to show that they are nilpotent and write the
Hamiltonian as quadratic forms of the supercharges. Next, we study the
interplay with other symmetries such as parity and spin reversal, and
establish the quadruplet structure of the eigenstates in special
momentum sectors.

In the sequel, we will frequently deal with operators that change the
number of sites of the chain. Hence, we will indicate by a subscript
the length of the chain the corresponding operator acts on. For
example, we denote by $Q_N$ the supercharge acting on the Hilbert
space for chains with $N$ sites etc.

\subsection{Elementary symmetries}
\label{sec:notations}

We start with a bit of notation. We denote by $\mathcal H_N=({\mathbb
  C}^2)^{\otimes N}$ the usual Hilbert space for $N$ spin-$1/2$
  particles in a chain. We will use the standard orthonormal basis in which the operators $\sigma^z_j$ are diagonal ($\sigma^x,\sigma^y,\sigma^z$ are the usual Pauli matrices). Its basis vectors are labeled by configurations $\alpha=\alpha_1\alpha_2\cdots\alpha_N$ with $\alpha_j = \,+$ (spin up) or $-$ (spin down) for the $j$-th spin, and the $\sigma_j^z$ operator acts according to
\begin{equation}
  \sigma_j^z|\alpha_1\cdots \alpha_j \cdots \alpha_N\rangle = \alpha_j|\alpha_1\cdots \alpha_j \cdots \alpha_N\rangle.
\end{equation}

Let us indicate here some elementary symmetries of the Hamiltonian \eqref{eqn:xyzham}, valid for any choice of $J_x,J_y,J_z$.
First, it is invariant under translation. We will often use the translation operator $T_N$ acting on basis states of $\mathcal H_N$ according to the usual rule
\begin{equation*}
   T_N|\alpha_1\cdots \alpha_{N-1}\alpha_N\rangle = |\alpha_N\alpha_1\cdots \alpha_{N-1}\rangle.
\end{equation*}
The translation invariance of the Hamiltonian implies that it commutes with this operator, and therefore they can be diagonalised simultaneously. Consider a state $|\psi\rangle$ such that $T_N|\psi\rangle =t_N |\psi\rangle$, then cyclicity implies $(T_N)^N=1$. Therefore the eigenvalue $t_N$ is an $N$-th  root of unity. Writing $t_N=e^{\i k}$ we see that the momentum $k$ has to be an integer multiple of $2\pi/N$. In this work, we focus mainly on momentum $k=0$ for chains of odd length, and momentum $k=\pi$ for chains of even length.

Moreover, the Hamiltonian is invariant under reversal of the order of all spins. We thus define a parity operation $P_N$ through
\begin{equation*}
  P_N |\alpha_1\alpha_2\cdots \alpha_{N-1}\alpha_N\rangle = |\alpha_N\alpha_{N-1}\cdots \alpha_2\alpha_1\rangle.
\end{equation*}
Obviously, we have $P_N^2=1$ and therefore eigenvalues $\pm 1$.

Finally, notice that the Hamiltonian is a quadratic form of the Pauli matrices. Therefore it remains unchanged under global rotations by an angle $\pi$ around any of the $x-,y-$ or $z-$axis. Let us first consider the $z$ axis. The rotation is given by $\i^N S_N$ where
\begin{equation}
  S_N = \sigma_1^z\sigma_2^z\cdots\sigma_N^z =\exp \frac{\i \pi}{2} \sum_{j=1}^N(1-\sigma_j^z).
  \label{eqn:sop}
\end{equation}
The right-hand side makes evident that $S_N$ has the eigenvalue $\pm 1$ on configurations with an even/odd number of spins $-$.

Considering instead rotations by the angle $\pi$ about the $x$-axis leads to the conclusion that the Hamiltonian commutes with the spin reversal operator $R_N$ defined through
\begin{equation}
  R_N = \sigma_1^x\sigma_2^x\cdots \sigma_N^x.
  \label{eqn:spinreversal}
\end{equation}
The operators $S_N$ and $R_N$ have the following (anti-)commutation relation:
\begin{equation}
  S_N R_N=(-1)^N R_N S_N \label{eqn:ac_sr}.
\end{equation}  
This implies in particular that for odd $N$ the spin-reversal operator couples the sectors with even and odd number of spins down. Therefore any eigenvalue of $H_N$ has even degeneracy and is at least doubly degenerate.

\subsection{Supercharges}
\label{sec:defsusy}

We proceed with the definition of the supercharges.
It is useful to start by recalling the usual ${\cal N}=2$
supersymmetry algebra \cite{witten:82}.  It is built from two
conjugate supercharges $Q, \,Q^\dagger$ that are nilpotent, i.e.\
their squares are zero: $Q^2=(Q^\dagger)^2=0$. A further symmetry
generator is the fermion number $F$. The supercharges obey the
relations $[F,Q]=Q$ and $[F,Q^\dagger]=-Q^\dagger$. Hence $Q$
increases the fermion number by one, while $Q^\dagger$ decreases
it. The Hamiltonian is given as anticommutator $H=\{Q,Q^\dagger\}$. It
conserves the fermion number and commutes with $Q$ and
$Q^\dagger$. The fact that $H$ is of this form implies that its
eigenvalues are non-negative. Any states with zero energy are
automatically ground states, and annihilated by both
supercharges. Thus they each are a supersymmetry singlet. Conversely, all
states with positive energy $E>0$ form doublets of the supersymmetry
algebra, with the fermion numbers of the two states in a doublet
differing by one.

We here define the analogous supercharges on the lattice. We start
with a generalisation of the work \cite{yang:04} and introduce
operators $q_j$ which map $\mathcal H_N$ to $\mathcal H_{N+1}$ through
action on site $j$ according to the following rules (the subscripts
denote the positions of the corresponding spins):
\begin{subequations}
\begin{align}
  q_j|\alpha_1\cdots \alpha_{j-1}\underset{j}{+} \alpha_{j+1}\cdots \alpha_N\rangle &=0, \\
   q_j|\alpha_1\cdots \alpha_{j-1}\underset{j}{-}\alpha_{j+1} \cdots \alpha_N\rangle &=(-1)^{j-1}\Bigl(| \alpha_1 \cdots \alpha_{j-1}\underset{j}{+}\underset{j+1}{+}\alpha_{j+1} \cdots \alpha_N\rangle \nonumber \\
   & \hspace{2cm} -\zeta|\alpha_1 \cdots \alpha_{j-1}\underset{j}{-}\underset{j+1}{-}\alpha_{j+1} \cdots \alpha_N\rangle\Bigr).
\end{align}
We see that while states with spins $+$ at site $j$ are annihilated by
$q_j$, the spins $-$ are transformed into pairs $++$ or $--$ with weight
$1$ and $-\zeta$ respectively. The pair creation implies a shift of
the spin sequence $\alpha_{j+1}\cdots \alpha_N$ by one site to the
right. This includes a ``string'' $(-1)^{j-1}$ which is
crucial in the following. 
To respect the periodic boundary conditions here, we need to build
eigenstates of the translation operator.  We thus introduce an
operator that creates a pair of like spins $++$ or $--$ on sites $N+1$
and $1$, namely
\begin{align}
  q_0|\alpha_1\cdots \alpha_{N-1}\underset{N}{+}\rangle &= 0,\\
   q_0|\alpha_1\cdots \alpha_{N-1}\underset{N}{-} \rangle  &=   
-\Bigl(|\underset{1}+\alpha_1\cdots \alpha_{N-1}\underset{N+1}+ 
\rangle-\zeta|\underset{1}-\alpha_1 \cdots \alpha_{N-1}\underset{N+1}- 
\rangle\Bigr).
\end{align}
\label{eqn:defq}
\end{subequations}
The operator $q_0$ acts always on the last site irrespectively of
length of the chain with no string attached.  

With the help of these
``local'' operators $q_j$ we construct a supercharge $Q_N$ in the following
way: suppose that $|\psi\rangle$ is an eigenstate of the translation
operator $T_N$ with some eigenvalue $t_N$. Then we define $Q_N|\psi\rangle =0$
unless $t_N=(-1)^{N+1}$. If $t_N=(-1)^{N+1}$ however, we define
\begin{equation*}
  Q_{N}=\left(\frac{N}{N+1}\right)^{1/2}\sum_{j=0}^N q_j.
\end{equation*}
For $\zeta=0$ these operators decrease the number of $-$ spins by one
but map a state of definite total magnetisation to another state of
definite total magnetisation \cite{yang:04}. For finite $\zeta$
however, this is not the case because the two states on the right-hand
side of \eqref{eqn:defq} differ in magnetisation by two. This is
related to the fact that the XYZ Hamiltonian can flip pairs $++$ and
$--$ of adjacent spins, and thus conserves magnetisation only mod 2.

Because of $q_0$ the operator $Q_N$ seems to distinguish the last site of the chain from the others. However, translation invariance removes this distinction, and we claim that $Q_N$ is
a well-defined mapping between the momentum spaces with $t_N=(-1)^{N+1}$
. 
To show this it is useful to understand the transformation properties of $q_j$ under translation. From their definition it is not difficult to show that
\begin{align}
  T_{N+1}q_jT_{N}^{-1}&=-q_{j+1},\quad j=0,\dots,N-1, \nonumber\\
  T_{N+1}q_N &= (-1)^N q_0. \label{eqn:trsl}
\end{align}
Consider now an eigenvector $|\psi\rangle$ of the translation operator for the chain with $N$ sites: $T_N|\psi\rangle = t_N|\psi\rangle,\,t_N=(-1)^{N+1}$. Upon action with $Q_N$ we produce a vector $|\phi\rangle = Q_{N}|\psi\rangle$ belonging to $\mathcal H_{N+1}$. The application of $T_{N+1}$ to this new vector leads to
\begin{align*}
  T_{N+1}|\phi\rangle &= \left(\sum_{j=1}^{N-1} T_{N+1}q_j T_N^{-1}+T_{N+1}q_N T_N^{-1} + T_{N+1}q_0 T_N^{-1}\right)T_N|\psi\rangle\\
  &= t_N \left(-\sum_{j=1}^N q_j +(-1)^N t^{-1}_Nq_0\right)|\psi\rangle.
\end{align*}
As $t_N=(-1)^{N+1}$ the new vector is an eigenvector of $T_{N+1}$, with eigenvalue $-t_N=t_{N+1}$, what proves our claim. Thus, we have $T_{N+1} Q_N T_N^{-1}=-Q_N$.

The fact that the operators $Q_N$ are \textit{bona fide} mappings between the momentum spaces of interest is crucial to the supersymmetric structure of the XYZ chain \eqref{eqn:xyzham} that we describe now. First, the supercharges have ``square zero'' in the sense that
\begin{equation}
  Q_{N+1}Q_N =0.
  \label{eqn:np}
\end{equation}
Hence they can be thought of as fermionic. Second, \textit{if restricted to the subsectors with translation eigenvalue $t_N=(-1)^{N+1}$} the XYZ-Hamiltonian can be constructed from the supercharges and their Hermitian conjugates. The latter are defined in the usual way: if $|\psi\rangle$ is a vector in $\mathcal H_N$, and $|\phi\rangle$ in $\mathcal H_{N+1}$ then $\langle \psi |Q_N^\dagger |\phi\rangle=\langle \phi|Q_N |\psi\rangle^\ast$. With this definition, the Hamiltonian can be written as an ``anticommutator''
\begin{equation}
  H_N= Q_{N-1}Q_{N-1}^\dagger+Q_N^\dagger Q_N.
  \label{eqn:xyzhamac}
\end{equation}
The proofs of \eqref{eqn:np} and \eqref{eqn:xyzhamac} are elementary
but cumbersome. We present the details in the appendices
\ref{app:nilpotency} and \ref{app:hamiltonian}. Here, we study their
consequences for the eigenvalue spectrum, and thus give an explanation
of the common eigenvalues for systems of different size. We will see
that all these properties are familiar from the theory of $\mathcal
N=2$ supersymmetric quantum mechanics \cite{witten:82}. 

First of all, consider the eigenvalue equation $H_N|\psi\rangle = E|\psi\rangle$. Projecting back on $|\psi\rangle$ we find
\begin{equation}
  ||Q_{N-1}^\dagger |\psi\rangle ||^2+||Q_{N}|\psi\rangle ||^2=E|||\psi\rangle ||^2.
  \label{eqn:norm}
\end{equation}
It follows that the spectrum is non-negative: all $E \geq 0$. Let us
first concentrate on strictly positive energies $E>0$. For a chain
with $N$ sites these energies come in pairs in the sense that a given
positive eigenvalue occurs in the spectrum at either $N+1$ or $N-1$
sites. This can be seen as follows. The structure of the Hamiltonian
in \eqref{eqn:np} and \eqref{eqn:xyzhamac} results in the commutation relation
\begin{equation}
  H_{N+1}Q_N - Q_N H_N =0.
  \label{eqn:crhq}
\end{equation}
Hence, if $|\psi\rangle$ is an eigenvector of $H_N$ with eigenvalue $E$ in the subspace with $t_N=(-1)^{N+1}$, then $Q_N|\psi\rangle$ is either zero or an eigenvector of $H_{N+1}$ with the same eigenvalue $E$. Likewise $Q_{N-1}^\dagger|\psi\rangle$ is either zero or an eigenvector of $H_{N-1}$ with the same eigenvalue $E$. However, one of the two vectors $Q_N|\psi\rangle$, $Q_{N-1}^\dagger|\psi\rangle$ must vanish\footnote{To show this we write $H_N = H_N^{(1)}+H_N^{(2)}$ with $H_N^{(1)}=Q_N^\dagger Q_N $ and $H_N^{(2)}=Q_{N-1}Q_{N-1}^\dagger$. Then $H_N^{(1)}H_N^{(2)}=H_N^{(2)}H_N^{(1)}=0$. Therefore, their respective eigenspaces associated with non-zero eigenvalues are orthogonal. If $H^{(1)}_N|\psi\rangle = E|\psi\rangle$ for some $E>0$ then $H^{(2)}_N|\psi\rangle=0$, by reprojection on $|\psi\rangle$ we  find $||Q_{N-1}^\dagger |\psi\rangle||^2=0$, and therefore $Q_{N-1}^\dagger |\psi\rangle=0$. It follows that $Q_N|\psi\rangle$ is non-zero because otherwise $|\psi\rangle$ would be a zero-energy state. Thus we found a doublet $(|\psi\rangle, Q_N|\psi\rangle)$. A similar argument applies to the eigenstates of $H^{(2)}_N$, and leads to pairs $(|\psi\rangle, Q_{N-1}^\dagger|\psi\rangle)$. Finally, as $H_N$ commutes with $H_N^{(1)}$ and $H_N^{(2)}$ it follows that all non-zero eigenstates of our Hamiltonian organise in pairs in the sense stated above.}. Hence, every eigenstate with non-zero energy is part of a doublet
\begin{equation*}
  (|\psi\rangle, Q_{N}|\psi\rangle) \quad \text{and so} \quad (|\psi\rangle, Q^\dagger_{N-1}|\psi\rangle).
\end{equation*}

Conversely, zero-energy states are unpaired (singlets). From \eqref{eqn:norm} it follows that they must be solution to the two equations
\begin{equation}
  Q_N|\psi\rangle =0, \quad Q_{N-1}^\dagger|\psi\rangle =0.
  \label{eqn:nullstates}
\end{equation}
It is well known that the solutions to these equations are related to the cohomology $\mathfrak H_N= \ker Q_{N}/\mathop{\text{im}} Q_{N-1}$ of the operator $Q_N$. Every zero-energy eigenstate corresponds to a non-trivial (non-zero) element in $\mathfrak H_N$. This can be seen indirectly: suppose that there are two linearly independent zero-energy states $|\psi_1\rangle$ and $|\psi_2\rangle$, and assume that they are in the same cohomology class. This means that there is a state $|\phi\rangle$ for the chain with $N-1$ sites such that
\begin{equation*}
  |\psi_1\rangle = |\psi_2\rangle+Q_{N-1}|\phi\rangle.
\end{equation*}
We act with $Q_{N-1}^\dagger$ on both sides, and apply \eqref{eqn:nullstates}. This yields $Q_{N-1}^\dagger Q_{N-1}|\phi\rangle=0$, and by reprojection on $|\phi\rangle$ to $||Q_{N-1}|\phi\rangle||^2=0$. This implies $Q_{N-1}|\phi\rangle=0$ and therefore $|\psi_1\rangle = |\psi_2\rangle$ -- in contradiction to the assumption of linear independence. Hence every non-trivial cohomology class contains exactly one ground state. This has two consequences. First, every eigenstate with non-zero energy that is annihilated by $Q_N$ can be written in the form $|\psi\rangle = Q_{N-1}|\phi\rangle$ where $|\phi\rangle$ is an eigenstate for the chain with $N-1$ sites. We will frequently use this property in the next section. Second, the number of zero energy ground states is the number of distinct non-trivial elements in the cohomology. We formulate the following 
\conjecture{For odd $N$ there are two non-trivial elements in $\mathfrak H_N$, whereas for even $N$ there are none.}

\medskip
 
We checked this conjecture up to $N=11$ sites by evaluation of the row and column ranks of the rectangular matrices $Q_N$. Notice that for $N$ odd it implies Stroganov's conjecture \cite{stroganov:01} on the existence of two zero energy ground states, provided that one can prove that they occur in the zero-momentum sector.

Before proceeding, let us make the following comment: both the nilpotency and the supersymmetric structure of the XYZ-Hamiltonian studied in this section are only valid in certain momentum sectors. The nature of their derivation, given in the appendices \ref{app:nilpotency} and \ref{app:hamiltonian}, reveals that this restriction comes from matching the periodic boundary conditions. Everything else follows from local relations. Thus, we conclude that the supersymmetry has to be present in the full problem for chains of infinite length.  

\subsection{Spin-reversal symmetry}

In this and the next section we provide a detailed discussion of the relation between supersymmetry and the other symmetries of the Hamiltonian introduced in section \ref{sec:notations}. Let us give a motivation for this. The relation $Q_{N+1} Q_N=0$ implies that we cannot relate chains with $N$ sites to $N+2$ sites by sole use of the supercharges defined in the previous sections.
Yet, a detailed inspection of the spectra for small system sizes suggests such a connection. The most simple example is the the eigenvalue $3+\zeta^2$, appearing once in the spectrum for $N=2$ sites, twice for $N=3$ sites and once for $N=4$ sites. The inspection of other non-zero eigenvalues reveals similar patterns. This hints at a larger symmetry algebra which we analyse in this section.

The existence of these degeneracies can be explained through a remarkably simple observation. The supercharges $Q_N$ introduced in section \ref{sec:defsusy} treat spins up and down in a very asymmetric way. However, the Hamiltonian commutes with the spin-reversal operator \eqref{eqn:spinreversal}, as pointed out in section \ref{sec:notations}. Therefore it seems natural to introduce the spin-reversed version of $Q_N$:
\begin{equation*}
   \widetilde Q_N = R_{N+1}Q_N R_N.
\end{equation*}
In order to understand the implications of this operator on the spectrum we need to work out the algebra generated $Q_N,\,\widetilde Q_N$ and their adjoints. The full list of relations reads
\begin{subequations}
\begin{align}
  & Q_{N}Q_{N-1} = \widetilde Q_{N}\widetilde Q_{N-1} =0, & \quad Q_{N-1}^\dagger Q_{N}^\dagger =  \widetilde Q_{N-1}^\dagger \widetilde Q_{N}^\dagger =0, \label{eqn:ac1}\\
   & \widetilde Q_{N}^\dagger Q_{N}+ Q_{N-1} \widetilde Q_{N-1}^\dagger=0, & Q_{N}^\dagger \widetilde Q_{N}+  \widetilde Q_{N-1} Q_{N-1}^\dagger=0 \label{eqn:ac2},\\
    &  \widetilde Q_{N} Q_{N-1} + Q_{N}\widetilde Q_{N-1} =0, &  Q_{N-1}^\dagger \widetilde Q_{N}^\dagger + \widetilde Q_{N-1}^\dagger Q_{N}^\dagger=0 \label{eqn:ac3}.
\end{align}
The Hamiltonian can be written as anticommutator of either set of supercharges
\begin{equation}
  H_N = Q_N^\dagger Q_N + Q_{N-1}Q_{N-1}^\dagger = \widetilde Q_N^\dagger \widetilde Q_N + \widetilde Q_{N-1}\widetilde Q_{N-1}^\dagger.
  \label{eqn:ac4}
\end{equation}
\label{eqn:n22susy}
\end{subequations}
In fact, even any linear combination $\alpha Q_N + \beta \widetilde Q_N$ with $|\alpha|^2 +|\beta|^2=1$ is an admissible supercharge that will generate the Hamiltonian.
We omit the proofs of \eqref{eqn:ac2} and \eqref{eqn:ac3} as they are tedious and very similar to the proof of nilpotency outlined in appendix \ref{app:nilpotency}.

Instead, we point out the striking analogies between \eqref{eqn:n22susy} and the $\mathcal N=(2,2)$ supersymmetry algebra in two dimensional quantum field theory (see for example \cite{hori:03}, chapter 22). The latter consists of \text{four} supercharges $\mathsf Q_\pm,\, \overline {\mathsf Q}_\pm$, Hamiltonian $\mathsf H$, momentum $\mathsf P$ and a fermion number $\mathsf F$. The algebra is defined through the relations
  \begin{subequations}
  \begin{align}
    & \mathsf Q_\pm^2 =\overline {\mathsf Q}_\pm^2 = 0,  \\
    & \{\mathsf Q_\pm ,\overline {\mathsf Q}_\mp \} = 0, \\
    & \{\mathsf{Q}_+,\overline {\mathsf Q}_+\}=\Delta,\quad  \{\mathsf{Q}_-,\overline {\mathsf Q}_-\}=\Delta^\ast ,\\
    & \{\mathsf Q_+,\mathsf Q_-\}=\mathsf H+ \mathsf P,\quad   \{\overline {\mathsf Q}_+,\overline {\mathsf Q}_-\}=\mathsf H- \mathsf P,\\
    &  [\mathsf F,\mathsf Q_\pm]=\pm \mathsf Q_\pm,\quad   [\mathsf F,\overline {\mathsf Q}_\pm]=\mp \overline {\mathsf Q}_\pm .
  \end{align}
  \label{eqn:n22qftsusy}
  \end{subequations}
together with the conjugation relations $\mathsf Q_\pm^\dagger = \mathsf Q_\mp$ and  $\overline{\mathsf Q}_\pm^\dagger = \overline {\mathsf Q}_\mp$. The definition $\mathsf H$ and $\mathsf P$ implies that they commute with all supercharges and the fermion number. The operators $\Delta$ and $\Delta^\ast$ are central elements. Non-zero values of the latter are usually an indication of topological sectors \cite{witten:78} which occur generically in non-compact spaces.

The similarity between the two algebraic structures
\eqref{eqn:n22susy} and \eqref{eqn:n22qftsusy} is certainly not a
coincidence. In fact, as mentioned in the introduction, the scaling
limit of the lattice model is described by the sine-Gordon field
theory at the supersymmetric point. Precisely, the limit is $N\to
\infty,\, \zeta \to 0$ with $g = \zeta N^{1/3}$ finite, and it yields
the sine-Gordon theory with bare mass $g$ and coupling $\beta =
\sqrt{16\pi/3}$ in conventional units. It was shown in
\cite{bernard:90} that the sine-Gordon field theory possesses an
affine quantum group symmetry $U_q(\widehat{\text{sl}}(2))$,
$q=\exp(-8\i\pi^2/\beta^2)$ with zero centre. For $\beta =
\sqrt{16\pi/3}$ the latter is known to
contain the $\mathcal N=(2,2)$ supersymmetry algebra \cite{bernard:90_2}, which indeed corresponds to our coupling.

Let us establish a dictionary between the lattice and field theory quantities. The first three lines of \eqref{eqn:n22susy} and \eqref{eqn:n22qftsusy} are in one-to-one correspondence, provided that we identify the field theory charges with some linear combinations of the lattice supercharges, and replace anticommutators by appropriate (and natural) expressions for the lattice operators. We see that there is no equivalent of the central charges $\Delta,\,\Delta^\ast$ in the lattice theory as the corresponding anticommutators are zero. We attribute this to the fact, that we have a finite, discrete and periodic space, and thus no topological/solitonic sector. The fourth line of \eqref{eqn:n22qftsusy} defines the field theory Hamiltonian and momentum. Whereas the former is related to the lattice XYZ Hamiltonian in a direct way, the identification between momenta in field and lattice theory needs to be supplemented by a shift by $(N+1)\pi \mod 2\pi$. The need for a shift becomes even more plausible when taking into account that the lattice equivalent of fermion number appears to be the number of sites of the spin chain. This is consistent with the Jordan-Wigner transformation appearing in the analysis of continuum limit of the XYZ chain \cite{luther:76}.

It is known that all non-zero energy states of a theory possessing the
$\mathcal N=(2,2)$ supersymmetry \eqref{eqn:n22qftsusy} are organised in quadruplets of the form $(|\psi\rangle; \mathsf Q_+|\psi\rangle, \overline {\mathsf Q}_-|\psi\rangle; \mathsf Q_+\overline {\mathsf Q}_-|\psi\rangle)$ with the state $|\psi\rangle$ being annihilated by $\mathsf Q_-$ and $\overline {\mathsf Q}_+$. The states in such a supermultiplet have all the same energy, and momentum, but differ in their fermion number. Given the similarities between the field theory and lattice algebra, it is natural to ask if there is a quadruplet structure in the lattice model. Indeed, we now show that the relations \eqref{eqn:n22susy} imply that it exists, at least in the momentum sectors considered in this paper. From the last section, we already know that the eigenstates $|\psi\rangle$ of the lattice Hamiltonian $H_N$ are part of doublets $(|\psi\rangle, Q_{N-1}^\dagger|\psi\rangle)$ or $(|\psi\rangle, Q_N|\psi\rangle)$. Without loss of generality, we focus on the second case $(|\psi\rangle, Q_N|\psi\rangle)$. Let us consider the vector $\widetilde Q_N|\psi\rangle$. There are two possibilities: it may either be non-zero or zero. First, suppose that $\widetilde Q_N|\psi\rangle$ is a non-zero vector. Clearly, it has the same energy as $Q_N|\psi\rangle$, and we might wonder if they coincide. In fact, we show that this cannot be the case, and that they are rather linearly independent. If there is linear dependence, then there must be non-zero numbers $\lambda$ and $\mu$ such that $\lambda Q_N|\psi\rangle + \mu \widetilde Q_N|\psi\rangle=0 $.
We show that $\lambda=0$ by applying $Q_N^\dagger$ from the left. For
the first term, we use $Q_N^\dagger Q_N|\psi\rangle =
H_N|\psi\rangle=E|\psi\rangle$; for the second term, we make use of
the anticommutation relation \eqref{eqn:ac2} and write
$Q^\dagger_N\widetilde Q_N|\psi\rangle=-\widetilde Q_{N-1}
Q^\dagger_{N-1}|\psi\rangle =0$. Thus, we are left with $-\lambda
E|\psi\rangle=0$ but because of $E>0$ we must have
$\lambda=0$. Likewise, one shows that $\mu =0$. Thus, the two vectors
are linearly independent. This implies in particular that in addition
to $Q_{N-1}^\dagger|\psi\rangle=0$ we have the equation $\widetilde
Q_{N-1}^\dagger|\psi\rangle=0$. Next, we increase once more the system
size: consider the vector $Q_{N+1}\widetilde
Q_N|\psi\rangle=-\widetilde Q_{N+1} Q_N|\psi\rangle$. This vector is
non-zero as again can be shown by an indirect proof: if for example $Q_{N+1}\widetilde Q_N|\psi\rangle=0$ then there would be some vector $|\phi\rangle$ for the chain with $N$ sites such that $\widetilde Q_N |\psi\rangle = Q_N|\phi\rangle$. If we premultiply this relation by $Q_N^\dagger$ we find on the left-hand side $Q_N^\dagger\widetilde Q_N |\psi\rangle = -\widetilde Q_{N-1} Q_{N-1}^\dagger|\psi\rangle=0$ because of \eqref{eqn:ac2}. Thus, the right-hand side becomes $Q_N^\dagger Q_N|\phi\rangle=0$ what implies $Q_N|\phi\rangle=0$. Yet, this is in contradiction to $\widetilde Q_N|\psi\rangle \neq 0$, proving our claim. We cannot apply more supercharges in order to increase the length of the chain because the state $Q_{N+1}\widetilde Q_N |\psi\rangle$ is annihilated by both $Q_{N+2}$ (trivially), and $\widetilde Q_{N+2}$ (because of \eqref{eqn:ac3}). Thus, we have constructed a quadruplet of one state at $N$ sites, \textit{two} states at $N+1$ sites, and one state $N+2$ sites, all of them having the same energy $E$ with respect to the corresponding Hamiltonians:
\begin{equation*}
  \left(|\psi\rangle;\,Q_N|\psi\rangle ,\, \widetilde Q_N|\psi\rangle;\,Q_{N+1}\widetilde Q_N|\psi\rangle\right).
\end{equation*}

The preceding construction assumes that $\widetilde Q_N|\psi\rangle$
is non-zero. Let us now consider the second case $\widetilde
Q_N|\psi\rangle=0$. As $E>0$ this can only be the case if there is a
vector $|\phi\rangle$ for the chain with $N-1$ sites such that
$|\psi\rangle = \widetilde Q_{N-1}|\phi\rangle$. Consider now the
state $Q_{N-1}|\phi\rangle$. It cannot be zero: otherwise, we could
write $0 = \widetilde Q_{N}(Q_{N-1}|\phi\rangle)=- Q_{N}(\widetilde
Q_{N-1}|\phi\rangle=-Q_N|\psi\rangle$ what contradicts our assumptions
as we started from a doublet $(|\psi\rangle,
Q_N|\psi\rangle)$. Moreover, the state $Q_{N-1}|\phi\rangle$ is
linearly independent from $|\psi\rangle$, as follows from the same argument as above. We have $Q_N|\psi\rangle = Q_N\widetilde Q_{N-1}|\phi\rangle=-\widetilde Q_N Q_{N-1}|\phi\rangle$, and hence a similar quadruplet structure as before, this time however with one state at $N-1$ sites, \textit{two} states at $N$ sites, and one state $N+1$ sites:
\begin{equation}
  \left(|\phi\rangle;\,Q_{N-1}|\phi\rangle ,\, \widetilde Q_{N-1}|\phi\rangle;\,Q_{N}\widetilde Q_{N-1}|\phi\rangle\right), \quad \widetilde Q_{N-1}|\phi\rangle=|\psi\rangle.
  \label{eqn:quadruplet}
\end{equation}

Therefore, all states with non-zero energy must be part of a quadruplet. We see that our argument leads automatically to a degeneracy in the ``middle'' of such a quadruplet. In fact, there is a non-trivial conserved charge that maps between these two states. It is given by
\begin{equation*}
  C_N = \widetilde Q_N^\dagger Q_N = - Q_{N-1}\widetilde Q_{N-1}^\dagger.
\end{equation*}
The anticommutation relations imply that it commutes with the Hamiltonian and has square zero:
\begin{equation*}
  [H_N, C_N]=0,\quad \text{and}\quad C_N^2=0.
\end{equation*}
Moreover, its Hermitian conjugate is the ``spin-reversed'' operator $C_N^\dagger = R_N C_N R_N$. They have the character of fermionic ladder operators. Indeed, let us consider the quadruplet \eqref{eqn:quadruplet} containing two states $|\psi\rangle= Q_{N-1}|\phi\rangle$ and $|\widetilde\psi\rangle= \widetilde Q_{N-1}|\phi\rangle$ at $N$ sites. We find the following relations
\begin{align*}
  C_N| \psi\rangle=0,& \quad C_N|\widetilde \psi\rangle=-E|\psi\rangle,\\
  C_N^\dagger| \widetilde\psi\rangle=0,& \quad C_N^\dagger|\psi\rangle=-E|\widetilde \psi\rangle.
\end{align*}
The other two states in the quadruplet, $|\phi\rangle$ and $Q_{N}\widetilde Q_{N-1}|\phi\rangle$, are annihilated by the corresponding operators $C_{N-1}$ and $C_{N+1}$ and their Hermitian conjugates.
Thus, $(|\psi\rangle, |\widetilde\psi\rangle)$ can be thought of a
doublet inside the quadruplet \eqref{eqn:quadruplet}. 

Even though $C_N$ is a bilinear in the supercharges, it still can be
thought of as fermionic in the following sense.
The symmetry operator $S_N$ defined in section
\ref{sec:notations} anti-commutes with the fermion: $C_N
S_N + S_N C_N=0$, as can be shown using \eqref{eqn:ac_sr} and the fact
that $S_{N+1}Q_N + Q_N S_N = 0$. If we now suppose that the state
$|\psi\rangle$ is an eigenstate of $S_N$ with eigenvalue $s=\pm 1$
then the anticommutation relation tells us that $|\widetilde
\psi\rangle$ is also an eigenstate of $S_N$, however with eigenvalue
$-s$. Therefore, we see that the fermionic operators $C_N$ provide a
mapping between the sectors with odd and even number of spins down. For
chains of odd length, this connection is already established through
the spin-reversal operator, as explained in section
\ref{sec:notations}. Namely, the states
$(|\psi\rangle, |\widetilde\psi\rangle)$ can be mapped onto each other
through spin reversal, as the state $|\phi\rangle$ is an eigenstate of
the spin-reversal operator $R_{N-1}$ when $N$ is odd.
For chains of even length, however, the spin-reversal operator fails to connect the two vectors, whereas the operator $C_N$ does this independently of the number of sites.

\subsection{Parity symmetry}

In the last part of this section, we analyse the relation between the supercharges and the the parity operation.
Using the definition of the ``local'' supercharges \eqref{eqn:defq} we find the simple transformation laws
\begin{align*}
  P_{N+1}q_j = (-1)^{N+1}q_{N-j+1}P_N,\quad P_{N+1}q_0 = q_0 P_N T_N.
\end{align*}
Let us now consider a state $|\psi\rangle$ for a chain of $N$ sites with both definite translational behaviour $T_N|\psi\rangle=t_N|\psi\rangle$ and definite parity $P_N|\psi\rangle=p_N|\psi\rangle,\, p_N=\pm 1$. The parity operation reverses momentum, as can be seen from the relation $P_N T_N P_N = T_N^{-1}$. This implies that the translation eigenvalue $t_N$ must be solution to $t^2_N=1$. Obviously, this is compatible with $t_N=(-1)^{N+1}$. Applying our rules, we find
\begin{align*}
  P_{N+1}Q_{N}|\psi\rangle = (-1)^{N+1}p\left(\sum_{j=1}^N q_j +(-1)^{N+1}t_N\, q_0\right)|\psi\rangle= (-1)^{N+1}p_N Q_{N}|\psi\rangle.
\end{align*}
Thus, we find that (1) for $N$ odd, $t_N=1$ the action of $Q_{N}$ preserves parity and (2) for $N$ even, $t_N=-1$ the action of $Q_{N}$ reverses parity.

We studied the parity sectors by means of exact diagonalisation of the Hamiltonian up to $N=11$ sites. For odd $N$ we observed that the spectrum of the parity odd sector is contained in the parity even sector. Thus, we arrive at the following

\conjecture{For odd $N$ and zero momentum, the spectrum in the odd parity sector $p_N=-1$ is contained in the spectrum of the even parity sector $p_N=+1$.}

\section{Supersymmetry and the transfer matrix of the eight-vertex model}
\label{sec:bethe}
In this section, we consider the transfer matrix of the zero-field
eight-vertex model. The main tool here is the Bethe ansatz for
the transfer matrix established by Baxter \cite{baxter:73,baxter:73_2,
baxter:73_3}. We provide a derivation of the supersymmetry from
this point of view, generalising the result of \cite{fendley:03} from
the critical point to the entire supersymmetric line.

After some basic definitions in section \ref{sec:basic}, we provide a
brief review of the Bethe ansatz for the eight-vertex model in the
root-of-unity case in section \ref{sec:baxterbethe}, in particular
recalling the necessary change of basis of the Hilbert space. In
section \ref{sec:derivsusy} we establish the supersymmetry in the new
basis. To make contact with the supercharges defined in our previous
discussion, we have to transform back to the canonical spin basis what
is discussed in \ref{sec:qspinrep}. This leads to some new conjectures
on the nature of the zero-energy ground states of the XYZ chain of odd
length.

\subsection{Basic definitions}
\label{sec:basic}
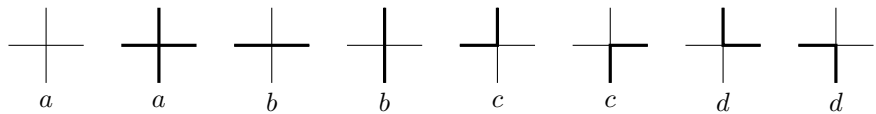
\begin{figure}[h]
  \centering
  \begin{tikzpicture}
    \foreach \x in {1,4,5.5,...,11.5}
    {
      \draw (-0.5+\x,0) -- (0.5 +\x,0);
      \draw (\x,-0.5) -- (\x,0.5);
    }
    
    \draw[very thick] (2,0)--(3,0);
    \draw[very thick] (2.5,0.5)--(2.5,-0.5);
    \draw[very thick] (3.5,0)--(4.5,0);
    \draw[very thick] (5.5,0.5)--(5.5,-0.5);
    \draw[very thick] (7,0.5)--(7,0)--(6.5,0);
    \draw[very thick] (9,0)--(8.5,0)--(8.5,-0.5);
	\draw[very thick] (10.5,0)--(10,0)--(10,0.5);
	\draw[very thick] (11,0)--(11.5,0)--(11.5,-0.5);
	
	\draw (1,-0.75) node {$a$}; \draw (2.5,-0.75) node {$a$};
	\draw (4,-0.75) node {$b$}; \draw (5.5,-0.75) node {$b$};
	\draw (7,-0.75) node {$c$}; \draw (8.5,-0.75) node {$c$};
	\draw (10,-0.75) node {$d$}; \draw (11.5,-0.75) node {$d$};

  \end{tikzpicture}
  \caption{The vertex configurations of the eight-vertex model (the
  bold edges have spin down). The associated weights $a,b,c,d$ are invariant under spin reversal in the zero field case.}
  \label{fig:configs}
\end{figure}

We start by recalling elementary facts about the eight-vertex model on
the square lattice \cite{baxterbook}. Each edge carries a classical ${\mathbb
  Z}_2$ ``spin'' variable $\pm$, corresponding to occupied/empty or spin up/down. The configurations are restricted in such a way that each vertex has an even number of spins down: the eight allowed vertex configurations are shown in figure \ref{fig:configs}.
We associate a Boltzmann weight to each vertex, and the weight of a
given lattice configuration is then simply the product over all the
vertex weights. In the ``zero-field'' case, the weights are invariant
under simultaneous reversal of all spins around a vertex. Thus, as
shown in figure \ref{fig:configs}, there are four distinct weights,
traditionally denoted by $a,b,c,d$. Suppose that the square lattice
has say $M$ rows and $N$ columns, and is wrapped around a torus (periodic boundary conditions along the two directions). Then the model can conveniently be studied by the row-to-row transfer matrix $\bm{\mathcal T}_N$, whose matrix elements are defined as the sum over all configurations along a horizontal line, compatible with the spin values on the vertical edges:
\begin{equation*}
  \langle \alpha'|\bm{\mathcal T}_N|\alpha\rangle = \sum_{\mu_1,\dots,\mu_N = \pm}
  \tikz[baseline]
  {
   \draw (0,0)--(2,0);
  \draw (0.5,-0.5)--(0.5,0.5);
  \draw (1.5,-0.5)--(1.5,0.5);
  \draw [dashed] (2,0)--(3,0);
  \draw (3,0)--(4,0);
  \draw (3.5,-0.5)--(3.5,0.5);
  \draw (0.,0) node[below] {\small $\mu_1$};
  \draw (1,0) node[below] {\small $\mu_2$};
  \draw (2,0) node[below] {\small $\mu_3$};
  \draw (3,0) node[below] {\small $\mu_N$};
  \draw (4,0) node[below] {\small $\mu_1$};
  \draw (0.5,-0.5) node[below]{\small $\alpha_1$};
\draw (1.5,-0.5) node[below]{\small $\alpha_2$};
\draw (3.5,-0.5) node[below]{\small $\alpha_N$};
\draw (0.5,0.5) node[above]{\small $\alpha_1'$};
\draw (1.5,0.5) node[above]{\small $\alpha_2'$};
\draw (3.5,0.5) node[above]{\small $\alpha_N'$};
  }.
\end{equation*}
The invariance of the vertex weights under spin reversal implies that $[\bm{\mathcal T}_N,R_N]=0$. Moreover, the vertex rule implies that the transfer matrix conserves the number of down spins mod $2$. Therefore we have $[\bm{\mathcal T}_N,S_N]=0$. Conservation of the total number of down spins is only possible in the six-vertex limit $d=0$ (or $c=0$).

To proceed we parametrise of the vertex weights in terms of Jacobi
theta functions, following the definitions of
\cite{bazhanov:05,bazhanov:06,mangazeev:10} and \cite{whittaker:27}:
\begin{align*}
  a =a(u)= \rho\, \vartheta_4(2\eta,q^2)\vartheta_4(u-\eta,q^2)\vartheta_1(u+\eta,q^2),\\
  b = b(u)=\rho\, \vartheta_4(2\eta,q^2)\vartheta_1(u-\eta,q^2)\vartheta_4(u+\eta,q^2),\\
  c = c(u)=\rho\, \vartheta_1(2\eta,q^2)\vartheta_4(u-\eta,q^2)\vartheta_4(u+\eta,q^2),\\
  d = d(u)= \rho\, \vartheta_1(2\eta,q^2)\vartheta_1(u-\eta,q^2)\vartheta_1(u+\eta,q^2).
\end{align*}
Here $u$ denotes the spectral parameter, $\eta$ the so-called crossing
parameter, and $q$ the elliptic nome. Moreover, we choose the overall
normalisation as $\rho=2/\vartheta_2(0,q)\vartheta_4(0,q^2)$. This choice ensures that
\begin{equation}
  h(u) = a(u)+b(u) = \vartheta_1(u,q),
  \label{eqn:defh}
\end{equation}
a function which we shall use quite often (the right-hand side follows from standard identities for Jacobi theta functions \cite{whittaker:27}). With this parametrisation two transfer matrices with different spectral parameters $u,u'$ commute:
\begin{equation}
  [\bm{\mathcal T}_N(u),\bm{\mathcal T}_N(u')]=0. \label{eqn:ctm}
\end{equation}
This implies that the series expansion of the transfer matrix in the spectral parameter around any point yields a family of commuting operators. The most simple ones are the translation operator $T_N$ and the XYZ-Hamiltonian
\begin{equation*}
T_N =h(2\eta)^{-N} \bm{\mathcal T}_N(\eta), \quad H_N = a(\eta)/b'(\eta)\bm{\mathcal T}_N(u)^{-1}\bm{\mathcal T}_N'(u)|_{u=\eta}.
\end{equation*}
The Hamiltonian reduces exactly to our problem \eqref{eqn:xyzham} if the crossing parameter is set to $\eta=\pi/3$. In this case, the variable $\zeta$ used to parametrise the supersymmetric line is related to the elliptic nome through
\begin{equation}
  \zeta = \left(\frac{\vartheta_1(2\pi/3,q^2)}{\vartheta_4(2\pi/3,q^2)}\right)^2.
  \label{eqn:defzeta}
\end{equation}

We are interested in using supersymmetry to study the eigenvalues and
eigenvectors of the eight-vertex model transfer matrix. Because of
\eqref{eqn:ctm} the eigenvectors do not depend on the spectral
parameter, and so coincide with those of the XYZ Hamiltonian $H_N$, up
to possible degeneracies. Such degeneracies do not seem to appear at
generic values of the crossing parameter, but only at the special
elliptic root of unity points $\eta = (m_1\pi+m_2\pi \tau)/L$, with
$m_1,m_2,L$ integers and $q=e^{\i \pi \tau}$, where additional
symmetries are present \cite{deguchi:02}. In our case $\eta=\pi/3$ we
have already shown that in the momentum sectors with $t_N =
(-1)^{N+1}$ the eigenvectors organise into singlets, or quadruplets
with the same value of $E$. Moreover, the eigenvectors in a given
quadruplet each can be labeled by a distinct quantum number; two of
them are for $N-1$ and $N+1$ sites, while we showed that the two at
$N$ sites have an even and odd numbers of spins down. Both the number
of sites and the number mod 2 of spins down are preserved by the
eight-vertex model transfer matrix, so barring any accidental
degeneracies, these correspond to distinct eigenvectors of the
transfer matrix as well.  Since $H_N$ is obtained from the logarithmic
derivative of $\bm {\mathcal T}_N(u)$, the fact that their
eigenvectors coincide makes it natural to hope that analogous
structure occurs in the spectrum of the transfer matrix. We here show
how at $\eta=\pi/3$ the supersymmetries described above indeed extend
to the transfer matrix, and so give relations among the eigenvalues.

As an indication of the special properties occurring at $\eta=\pi/3$,
we note that the zero-energy states $|\Psi^\pm\rangle$ of $H_N$ for odd $N$, i.e.\
the supersymmetry singlets, have very simple transfer-matrix
eigenvalues $\bm {\mathcal T}_N(u)|\Psi^\pm\rangle = \mathcal
T_N(u)|\Psi^\pm\rangle$. They are given by
\cite{stroganov:01,stroganov:01_2}
\begin{equation*}
  \mathcal T_N(u) = h(u)^N = \vartheta_1(u,q)^N.
\end{equation*}
The simplicity of this expression stresses the special nature of the
two eigenstates. In the sequel we will see that the study of the
transfer matrix eigenvalues leads naturally to a distinction of these
states from the other eigenvectors.

\subsection{Review of Baxter's Bethe ansatz}
\label{sec:baxterbethe} 

Here we summarise the aspects of the coordinate-type Bethe ansatz
\cite{baxter:73,baxter:73_2,baxter:73_3} relevant to our 
derivation of the supersymmetry in the eight-vertex
model at $\eta=\pi/3$. 

\paragraph{\it Path basis.} The transfer matrix of the eight-vertex model has no obvious particle-number conservation (such as conservation of the number of down-spins). This is a central difficulty when compared to the six-vertex model. In \cite{baxter:73_2} Baxter developed a way to overcome this problem through the introduction of a basis upon which the transfer matrix acts in a way that resembles the six-vertex case.

For $N$ sites the new basis vectors are labeled by a sequence of integers
$\ell_1,\ell_2,\dots,\ell_N,\ell_{N+1}$ such that
$|\ell_{j+1}-\ell_j|=1,\,j=1,\dots,N$. It is useful to think of a path
starting at some height $\ell_1=\ell$ with the restriction that
consecutive heights differ by $\pm 1$. In the following, we will
therefore frequently call the corresponding set of vectors in
$\mathcal H_N$ the ``path basis''. A down step or
particle occurs at site $j$ if $\ell_{j+1}-\ell_j=-1$, and an up step occurs
otherwise. The path is completely characterised by $\ell$ and the
positions $x_1,\dots,x_m$ of its $m$ down steps. Hence for
$x_{k}<j<x_{k+1}$ the local heights are given by
\begin{equation}
  \ell_j = \ell+j-(2k+1).
  \label{eqn:heightvalue}
\end{equation}
The basis vectors are given as an $N$-fold tensor product of local vectors $\Phi_{\ell,\ell'}$ in $\mathbb C^2$: 
\begin{equation*}
  |\ell; x_1,\dots, x_m\rangle_N = \bigotimes_{j=1}^N |\Phi_{\ell_j,\ell_{j+1}}\rangle.
\end{equation*}
The factors are constructed from the local heights via
\begin{equation}
  \begin{array}{l}
  |\Phi_{\ell,\ell+1}\rangle =
    \vartheta_1(s+(2\ell+1)\eta,q^2)|+\rangle\ +\ 
    \vartheta_4(s+(2\ell+1)\eta,q^2)|-\rangle,\\
  |\Phi_{\ell+1,\ell}\rangle =
    \vartheta_1(t+(2\ell+1)\eta,q^2)|+\rangle\ +\ 
    \vartheta_4(t+(2\ell+1)\eta,q^2)|-\rangle,
  \end{array}
  \label{eqn:factors}
\end{equation}
where $|\pm\rangle$ are the local spin$-1/2$ basis vectors, and $s$ and $t$ arbitrary parameters such that the two vectors
are linearly independent. 

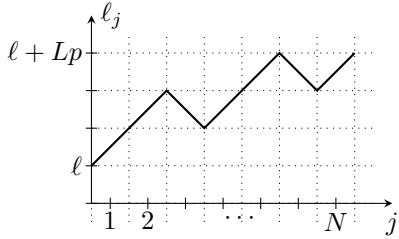
\begin{figure}[h]
  \centering
  \begin{tikzpicture}[>=stealth]
     \draw[scale=0.5,thick] (0,0)--(2,2)--(3,1)--(5,3)--(6,2)--(7,3);
           
     \draw[<->] (0,2) -- (0,-0.5) -- (4,-0.5);
     \foreach \x in {0.25,0.75,...,3.25}
       \draw[xshift=\x cm,yshift=-0.5cm] (0,-0.075)--(0,0.075);
     
    \foreach \y in {-0.5,0,...,1.5}
       \draw[yshift=\y cm] (-0.075,0)--(0.075,0);
 
    \draw (0.25,-0.5) node[below] {$1$};
    \draw (0.75,-0.5) node[below] {$2$};
    \draw (2,-0.5) node[below] {$\cdots$};
    \draw (3.25,-0.5) node[below] {$N$};
    \draw (4,-0.5) node[below] {$j$};
    \draw (0,0) node [left] {$\ell$};
    \draw (0,1.5) node [left] {$\ell+Lp$};
    \draw (0,2) node [right] {$\ell_j$};

     \clip[scale=0.5] (0,-1.5) rectangle (7.5,3.5);
     \draw[scale=0.5,dotted] (-1,-2) grid (8,5);

  \end{tikzpicture}

  \caption{A typical path for $L=3$, two particles and $p=1$.}
  \label{fig:path}
\end{figure}

The transfer matrix maps the set of these states onto themselves and conserves the number of particles, provided that the following condition is met \cite{baxter:73_2}: for generic values of the crossing parameter $\eta$ the initial and final height are identical $\ell_1=\ell_{N+1}$, and therefore $N=2m$.
In the special case of elliptic roots of unity $\eta = (m_1\pi+m_2\pi \tau)/L$ however, this constraint can be relaxed to
\begin{equation}
  N-2m = Lp,\quad \text{for some }p\in \mathbb Z
  \label{eqn:condrou}
\end{equation}
because of the periodicity of the Jacobi theta functions involved in the construction of the vectors.
The height difference between starting point and endpoint of the path is thus $Lp$ as shown in figure \ref{fig:path}. Moreover, because of the periodicity of the theta functions it is sufficient to restrict the initial height to $\ell_1=0,1,\dots,L-1$ in this case.

It is instructive to compute the maximal dimension $d_N$ of the subspace spanned by these vectors in the root-of unity-case by simple counting of the paths. There are ${N \choose m}$ arrangements of $m$ particles, provided that \eqref{eqn:condrou} holds. Let us introduce an indicator function $\delta_L(n)$ which is $1$ if $n=0 \mod L$, and $0$ otherwise. We have the convenient representation
\begin{equation*}
  \delta_L(n) = \frac{1}{L}\sum_{j=0}^{L-1}e^{2\pi\i  j n/L}.
\end{equation*}
We weight this by the number of arrangements and an additional factor $L$ which takes into account the different choices for $\ell =0,1,\dots, L-1$. Summation over $m$ yields
\begin{equation*}
  d_N = L\sum_{m=0}^N \delta_L(N-2m){N \choose m} =2^N\sum_{j=0}^{L-1}\left(\cos \left(\frac{2\pi j}{L}\right)\right)^N.
\end{equation*}
For the case of interest $L=3$, we find
\begin{equation*}
  d_N = 2^{N}+2(-1)^N.
\end{equation*}
We know that the dimension of the full Hilbert space $\mathcal H_N$ is $2^N$.
If we assume that all the vectors associated to paths are linearly independent then we conclude that for even $N$ the path basis is redundant. For odd $N$ however at least two vectors are missing, and thus the path basis does not span the entire Hilbert space. For small finite-size systems, it seems that exactly two vectors are missing, i.e.\ the existing $2^N-2$ vectors are linearly independent. We shall assume the linear independence in the following, and will conjecture later that the two-dimensional complement of the path basis at $N$ odd is spanned by the ground states of the XYZ Hamiltonian.

\paragraph{\it Eigenvectors and Bethe equations.} The next step consists of decomposing the eigenvectors of the transfer matrix in terms of vectors in the path basis:
\begin{equation}
  |\psi\rangle = \sum_{\ell=1}^{L}\omega^\ell\sum_{\{x_j\}}\psi(\ell;x_1,\dots, x_m)|\ell;x_1,\dots,x_m\rangle.
  \label{eqn:wavevector}
\end{equation}
The summation over the positions of the particles is carried out in an ordered way $1\leq x_1 <x_2<\cdots <x_m\leq N$.
Moreover $\omega$ is an $L$-th root of unity: $\omega^L=1$. The wave functions $\psi(\ell;x_1,\dots, x_m)$ are obtained through a Bethe-type ansatz. In order to describe it we need Baxter's ``single particle'' functions and ``wave vectors'' defined through
\begin{equation}
  g_j(\ell,x)= e^{\i k_j x} \frac{h(w_{\ell+x-1}-\eta-u_j)}{h(w_{\ell+x-2})h(w_{\ell+x-1})},\quad e^{\i k_j}=\frac{h(u_j+\eta)}{h(u_j-\eta)}.
  \label{eqn:defg}
\end{equation}
Here we used the function
\begin{equation}
  w_\ell=(s+t)/2-\pi/2+2\ell\eta \label{eqn:defw}
\end{equation}
which is linear in $\ell$, and contains the free parameters. The numbers $u_1,\dots,u_m$ are the \textit{Bethe roots} to be determined. With this notation the wave function is given in typical Bethe-ansatz form by
\begin{equation}
  \psi(\ell|x_1,\dots, x_m) = \sum_{\pi\in S_m} A_\pi g_{\pi(1)}(\ell,x_1)g_{\pi(2)}(\ell-2,x_2)\cdots g_{\pi(m)}(\ell-2(m-1),x_m).
  \label{eqn:defpsi}
\end{equation}
Here the sum runs over all permutations $\pi$ of $m$ objects. The $m!$ coefficients $A_\pi$ satisfy the following relation: if $\tau$ is a transposition exchanging $j$ and $j+1$, then we have for the permutation $\pi'=\pi\circ \tau$ the relation
\begin{equation}
  \frac{A_{\pi'}}{A_{\pi}} =  -\frac{h(u_{\pi(j+1)}-u_{\pi(j)}+2\eta)}{h(u_{\pi(j)}-u_{\pi(j+1)}-2\eta)}.
  \label{eqn:scattering}
\end{equation}
The left-hand side is commonly interpreted as the (bare) scattering
matrix between two particles with ``rapidities'' $u_{\pi(j)}$ and
$u_{\pi(j+1)}$. If they coincide then we find $A_{\pi'}=-A_\pi$,
implying that the Bethe wave function vanishes\footnote{
Let us suppose that two Bethe roots have the same value, say $u_{m-1}=u_m$.
Now we modify the sum over permutations in \eqref{eqn:defpsi} according to $\sum_{\pi} f_\pi = \sum_{\pi} f_{\tau\circ \pi}$ for some function $f$ on the symmetric group $S_m$, where $\tau$ is an arbitrary permutation of $m$ objects. We choose $\tau$ to be the transposition of $m-1$ and $m$. For any $\pi$ define pre-images $n_1,n_2$ according to $\pi(n_1)=m-1$ and $\pi(n_2)=m$, then we find
\begin{align*}
  \psi(\ell|x_1,\dots, x_m)=&\sum_{\pi} A_{\pi'}\prod_{j\neq n_1,n_2} g_{\pi(j)}(\ell-2(j-1),x_j)\\
   &\times g_{\pi(n_2)}(\ell-2(n_1-1),x_{n_1})g_{\pi(n_1)}(\ell-2(n_2-1),x_{n_2})
\end{align*}
According to our assumption we have $A_{\pi'}=-A_\pi$. Moreover, from $u_{m-1}=u_m$ and \eqref{eqn:defg}, it is not difficult to see that $g_{m-1}(\ell,x)= g_m(\ell,x)$.  Therefore we may write $g_{\pi(n_2)}(\ell-2(n_1-1),x_{n_1})g_{\pi(n_1)}(\ell-2(n_2-1),x_{n_2})=g_{\pi(n_1)}(\ell-2(n_1-1),x_{n_1})g_{\pi(n_2)}(\ell-2(n_2-1),x_{n_2})$. Using these facts, we see that
\begin{equation*}
\psi(\ell|x_1,\dots, x_m)=-\psi(\ell|x_1,\dots, x_m),
\end{equation*}
and therefore the wave function vanishes.}. This will be very
important in our analysis.

The Bethe roots $u_1, \dots, u_m$ remain to be determined. Baxter showed in \cite{baxter:73_3} that if they solve the Bethe equations
\begin{equation}
  \left(\frac{h(u_j+\eta)}{h(u_j-\eta)}\right)^N=-\omega^{2}\prod_{k=1}^m\frac{h(u_j-u_k+2\eta)}{h(u_j-u_k-2\eta)},
  \label{eqn:bae}
\end{equation}
then \eqref{eqn:wavevector} is an eigenvector of the transfer matrix $\bm{\mathcal T}_N(u)|\psi_N\rangle=\mathcal T_N(u)|\psi_N\rangle$. The corresponding eigenvalue can be obtained from the so-called $\mathcal T \mathcal Q$-equation
\begin{equation}
  \mathcal T_N(u)\mathcal Q_N(u) = \omega\phi_N(u-\eta)\mathcal Q_N(u+2\eta)+ \omega^{-1}\phi_N(u+\eta)\mathcal Q_N(u-2\eta),
  \label{eqn:tqequation}
\end{equation}
where $\phi_N(u) = h(u)^N = \vartheta_1(u,q)^N$, and
\begin{equation*}
  \mathcal Q_N(u) = \prod_{j=1}^m h(u-u_j)
\end{equation*}
is an elliptic polynomial with zeroes at the Bethe roots. In particular, setting $u=\eta$ we find that the eigenvalue $t_N$ of the translation operator $T_N$ is given in terms of the $\mathcal Q$-function as
\begin{equation}
  t_N = \omega^{-1}\frac{Q_N(-\eta)}{Q_N(\eta)} = \omega^{-1}\prod_{j=1}^m \frac{h(u_j+\eta)}{h(u_j-\eta)} =\omega^{-1}\prod_{j=1}^m e^{\i k_j},
  \label{eqn:trslev}
\end{equation}
where in the last step we used \eqref{eqn:defg}.

\subsection{Derivation of the supersymmetry from the Bethe ansatz}
\label{sec:derivsusy}
We now use the Bethe ansatz to establish for the case $\eta=\pi/3$ the
supersymmetry connecting systems with different numbers of sites $N$
and $N\pm 1$. 

We start by noting that from \eqref{eqn:condrou}, the number of
particles $m$ in the path when $L=3$ must obey
\begin{equation*}
  N-2m=3p
\end{equation*}
for some integer $p$. This relation is compatible with the
simultaneous replacement $N\to N'=N-j$, $m\to m'=m+j$ and $p\to
p'=p-j$ for some integer $j$. The supersymmetry charge $Q_N$ 
studied in section \ref{sec:hamiltonian} {increases} the number of sites
by one. However, in the context of the Bethe ansatz it turns out
particularly convenient to consider an action like that of $Q^\dagger_{N-1}$: we choose $j=1$ and therefore \textit{decrease} the length of the chain by one while adding a particle to the system. We discuss the relation with the supercharges studied previously in the next section.

Our strategy is to construct from a given solution $u_1,\dots, u_m$ of Bethe's equations at $N$ sites a new solution $\tilde {u}_1,\dots, \tilde u_m, \tilde u_{m+1}$ at $N-1$ sites. We shall verify that a solution to this problem is simply given by $\tilde u_j = u_j$ for $j<m+1$ and $u_{m+1}=\pi$. Indeed, for the smaller system the first $m$ Bethe equations with this choice become
\begin{equation*}
  \left(\frac{h(u_j+\eta)}{h(u_j-\eta)}\right)^{N-1}=-\omega^{2}\prod_{k=1}^m \frac{h(u_j-u_k+2\eta)}{h(u_j-u_k-2\eta)}\times \frac{h(u_j-\pi+2\eta)}{h(u_j-\pi-2\eta)}.
\end{equation*}
Using \eqref{eqn:bae} and the antiperiodicity $h(u+\pi)=-h(u)$, this equation reduces to
\begin{equation*}
  \frac{h(u_j-\eta)}{h(u_j+\eta)}=\frac{h(u_j-\pi+2\eta)}{h(u_j+\pi-2\eta)}.
\end{equation*}
It holds for generic $u_j$ if $\eta=\pi-2\eta \mod \pi$, and thus in particular for the value $\eta = \pi/3$ we are interested in. However, we still have to check the $(m+1)$-th Bethe equation. We find
\begin{equation*}
  (-1)^{N+1} = \omega^{2}\prod_{j=1}^m\frac{h(u_j+\eta)}{h(u_j-\eta)}.
\end{equation*}
On the right-hand side we recognise the eigenvalue of the translation operator for the system with $N$ sites \eqref{eqn:trslev}. We conclude that the operation is possible only if $t_N = (-1)^{N+1}\omega^3$. But recall from the last section that for $\eta = \pi m/L$ the number $\omega$ is an $L$-th root of unity, in our case thus $\omega^3=1$, and therefore we find the symmetry in the momentum sector with
\begin{equation}
  t_N=(-1)^{N+1}
  \label{eqn:tev}.
\end{equation}
Consistency thus requires a restriction to the momentum sectors studied in section \ref{sec:hamiltonian}.

As a side comment, let us notice that we could have started from the
$\mathcal T \mathcal Q$-equation with arbitrary $\omega$, deduced the
Bethe equations from the requirement that $\mathcal {T}_N(u)$ is an
entire function and \textit{imposed} the lattice supersymmetry. Asking
for consistency would have led us to $\omega^3=1$. In the six-vertex
limit where the working is completely analogous, write $\omega=e^{\i
  \phi}$ so that $\phi$ has the interpretation of a twist angle. We
conclude that the twists leading to the symmetry here are
$\phi=0,\pm 2\pi/3$, as follows from the observations of \cite{fendley:03}. These are precisely the values for which special ground state eigenvalues of the transfer matrix, as well as relations to alternating sign matrices, appear \cite{razumov:00,razumov:01,degier:02}.

As a next step, we determine the relation between the corresponding eigenvalues of the transfer matrix from \eqref{eqn:tqequation}. For the $\mathcal Q$-function we find
\begin{equation*}
  \mathcal Q_{N-1}(u) = \prod_{k=1}^{m} h(u-u_j)\times h(u-\pi)=-h(u) \mathcal Q_N(u).
\end{equation*}
Using this relation, we deduce that
\begin{equation}
  \mathcal T_N(u)+  h(u)\mathcal T_{N-1}(u)=0.
  \label{eqn:relt}
\end{equation}
Setting $u=\eta$ we obtain a relation between the eigenvalues of the translation operators for both systems $t_{N-1}=-t_N = (-1)^N$. This is consistent with \eqref{eqn:tev}. For odd $N$ we obtain thus the zero-momentum sector (invariant under translation), whereas for even $N$ it is the $\pi$-momentum sector. This fits well the picture suggested by \eqref{eqn:defg}: the $(m+1)$-th particle with $u_{m+1}=\pi$ has momentum $k_{m+1}=\pi$, and therefore the eigenvalue of the translation operator is changed by a sign. 

\paragraph{\it Relation between eigenvectors.} The preceding operation should manifest itself as an operation on the Hilbert space (or at least the special momentum sectors). In fact, we would like to introduce an operator $\hat Q_{N}:\mathcal H_N \to \mathcal H_{N+1}$ (not to be confused with Baxter's $\bm{\mathcal Q}$-matrix) such that the eigenvectors of the transfer matrix that can be obtained from the Bethe ansatz are related according to
\begin{equation*}
  |\psi_{N-1}\rangle = \hat Q_{N-1}^\dagger|\psi_N\rangle.
\end{equation*}
Twofold application of $\hat Q_N^\dagger$ would lead to the injection of two particles with momentum $\pi$. However, in this case the Bethe wave function vanishes. Hence we can write on the subspace spanned by the path basis
\begin{equation*}
  \hat Q_{N-1}^\dagger \hat Q_N^\dagger = 0, \quad \text{or} \quad \hat Q_N \hat Q_{N-1}=0.
\end{equation*}

\begin{figure}[h]
  \centering
  \begin{tikzpicture}[scale=0.5]
    \begin{scope}
    \draw (0,2)--(1,1)--(5,5)--(6,4);
    \draw[very thick] (2,2)--(4,4);
    \draw (0.5,-1) node {$x_{k-1}$};
    \draw (5.5,-1) node {$x_k$};
    \draw(2.5,-1) node {$x$};
    
    \draw (0,0)--(6,0);
    \foreach \x in {0,...,5}
      \filldraw(\x+0.5,-0.1) -- (\x+0.5,+0.1);
    \clip  (-0.5,0) rectangle (6.5,5.5);
    \draw[dotted] (-1,0) grid (7,6);
    \end{scope}

    \draw[->] (7,2.5)--(9,2.5);
    
    \begin{scope}
    \draw[xshift=10cm] (0,2)--(1,1)--(2,2)--(3,1)--(4,2)--(5,1);
    \draw[very thick,,xshift=10cm] (2,2)--(3,1);
      \draw[xshift=10cm] (0.5,-1) node {$x_{k-1}$};
    \draw[xshift=10cm] (4.5,-1) node {$x_k-1$};
    \draw[xshift=10cm] (2.5,-1) node {$x$};
    
     \draw[xshift=10cm] (0,0)--(5,0);
    \draw[dotted,xshift=10cm] (-1,0)--(0,0);
    \draw[dotted,xshift=10cm] (5,0)--(6,0);
    \foreach \x in {0,...,4}
      \filldraw[xshift=10cm] (\x+0.5,-0.1) -- (\x+0.5,+0.1);
    \clip[xshift=10cm]  (-0.5,0) rectangle (5.5,5.5);
    \draw[xshift=10cm,dotted] (-1,0) grid (6,6);
    \end{scope}

  \end{tikzpicture}
  \caption{Local action of $\hat Q_{N-1}^\dagger$: two steps up are transformed to a step down.}
  \label{fig:localtrsf}
\end{figure}
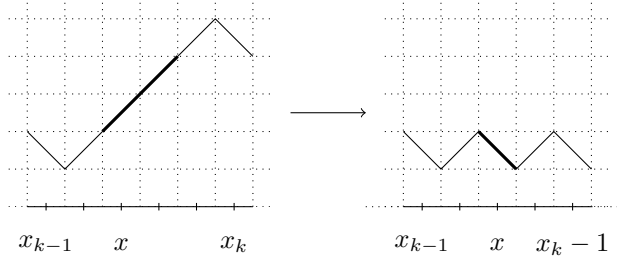

We will now derive the explicit form of these operators, starting from the definition of the wave functions \eqref{eqn:defpsi}. To manipulate them, we need an explicit expression for the amplitudes $A_\pi$. In fact, their defining equation \eqref{eqn:scattering} can be solved up to a factor:
\begin{equation}
  A_\pi = \sgn \pi \prod_{1\leq i< j\leq m} h(u_{\pi(i)}-u_{\pi(j)}+2\eta).
  \label{eqn:defamplitude}
\end{equation}
Let us consider the wave function \eqref{eqn:defpsi} for $m+1$ particles, one of them having momentum $u_{m+1}=\pi$. It is a sum over permutations $\pi$ of $\{1,2,\dots, m+1\}$. For a start, let us consider in this sum only the permutations with $\pi(r)=m+1$ for some fixed $r=1,\dots,m+1$. Any such permutation can be decomposed according to $\pi = \pi'\circ \pi''$ where $\pi'(m+1)=m+1$ and
\begin{equation*}
  \pi''(j) =
  \begin{cases}
    j, & j<r\\
    m+1, & j=r\\
    j-1, & j>r
  \end{cases}.
\end{equation*}
We have the signature $\sgn \pi'' = (-1)^{m+1-r}$ and thus $\sgn \pi = (-1)^{m+1-r}\sgn \pi'$. We use this in order to evaluate the Bethe amplitude \eqref{eqn:defamplitude} in terms of the permutation $\pi'$. After some algebra one finds
\begin{align*}
  A_\pi
  &= \sgn \pi' \prod_{1\leq i< j \leq m} h(u_{\pi'(i)}-u_{\pi'(j)}+2\eta) \prod_{i=1}^m h(u_{i}-\eta)\prod_{j=r+1}^{m+1} e^{\i k_{\pi(j)}}.
\end{align*}
We see that the only $r$-dependent term is the last product: a string of wave-vectors. Thus, we must understand how this affects the corresponding single-particle functions. For $j>r$ we notice the identity
\begin{equation*}
  e^{\i k_{\pi(j)}}g_{\pi(j)}(\ell - 2j+2,x_j)=g_{\pi'(j-1)}(\ell - 2(j-1)+2,x_j+1),
\end{equation*}
which holds only because of $\eta=\pi/3$. This is already enough to simplify the wave function. As $\pi'$ leaves $m+1$ unchanged, we can think of it as a permutation of only $m$ objects. This can of course be done for any value of $r$ separately. Collecting the different contributions, we find after some algebra a recursion relation for the wave functions
\begin{align*}
  \psi(\ell; x_1,\dots, x_{m+1})=&\prod_{i=1}^m h(u_{i}-\eta)\sum_{r=1}^{m+1}g_{m+1}(\ell-2(r-1),x_r)\nonumber\\
  & \qquad\times \psi(\ell; x_1,\dots,x_{r-1},x_{r+1}+1,\dots, x_{m+1}+1).
\end{align*}
The wave functions on the right hand side are the ones involving only the Bethe roots $u_1,\dots, u_m$, thus precisely the ones for the problem at $N$ sites. However, notice that the positions of the particles are not arbitrary: for the $r$-th term we have $x_{r+1}+1-x_{r-1}\geq 2$. The picture becomes a little more transparent if we consider the induced operation on the basis states\footnote{Strictly speaking, $\hat Q_{N-1}^\dagger$ only acts on momentum states with $t_N=(-1)^{N+1}$. Clearly, these can be written as a superposition of path states. Thus, for simplicity, we present the action of the supercharge on the vectors in this decomposition.}:
\begin{align}
  \hat Q_{N-1}^\dagger|\ell;x_1,\dots,x_{m}\rangle_N
  =\mathcal {C}\sum_{r=1}^{m+1}&\sum_{x=x_{r-1}+1}^{x_r-2}g_{m+1}(\ell-2(r-1),x) \nonumber\\
  & \times
  |\ell;x_1,\dots,x_{r-1},x,x_{r+1}-1,\dots ,x_{m}-1\rangle_{N-1}
  \label{eqn:defqhat}
\end{align}
where we set $x_0=0$; $\mathcal C$ is a normalisation constant which can be chosen arbitrarily.
We see that $\hat Q_{N-1}^\dagger$ inserts a new particle at position $x$ between existing ones at $x_{r-1}$ and $x_r$, provided that they are at least two sites apart (in order to guarantee $x_{r-1}<x<x_{r}-1$). 
More precisely the operation $\hat Q_{N-1}^\dagger$ transforms locally two consecutive up-steps at $(x,x+1)$ to a single down-step at $x$, while shifting all particles on its right by one step to the left as illustrated in figure \ref{fig:localtrsf}. 

This is weighted by $\mathcal{C} g_{m+1}(\ell-2k+2,x_k=x)$. Notice that the local height at $x+1$ is given by $\ell_x=\ell+{x+1}-2(k-1)$. Combining this with the definition of the single-particle wave function \eqref{eqn:defg}, we conclude that apart from the string the weight can be expressed in terms of the local height between the two up-segments alone (to see this recall that $h(u)$ is $2\pi$-periodic and $w_{x+3}=w_x+2\pi$). More explicitly, we make the convenient choice $\mathcal C = \prod_{\ell=1}^3{h(w_{\ell})}$ and find the local weight
\begin{equation}
  \mathcal C g_{m+1}(\ell-2k+2,x_k=x)=(-1)^x h(w_{\ell_{x+1}})^2.
  \label{eqn:weightqhat}
\end{equation}

This completes the definition of the supercharges acting on the path basis. From \eqref{eqn:relt}, we conclude that they act as intertwiners for the transfer matrices:
\begin{equation}
 \bm {\mathcal T}_{N}(u)\hat Q_{N-1}+ h(u)\hat Q_{N-1}\bm{\mathcal T}_{N-1}(u)=0,
 \label{eqn:intertwt}
\end{equation}
on the momentum sectors with $t_N = (-1)^{N+1}$.
Of course, this relation only holds on the subspace spanned by the path basis.

We finish this section by pointing out that -- as in the case of the
XYZ chain -- the notion of ``particle'' is somewhat arbitrary. One
could as well have chosen the steps up as particles. From the local
vectors \eqref{eqn:factors} we see that this corresponds essentially
to exchanging the parameters $s$ and $t$.  This would lead to a second
supersymmetry operation with the same local weights
\eqref{eqn:weightqhat}, which transforms locally two consecutive steps down
to a single step up, and thus resembling strongly the case studied in
section \ref{sec:hamiltonian}. We will discuss their connections in the next section.

\subsection{Supercharges in the spin representation and the XYZ ground states}
\label{sec:qspinrep}
Having found an operator $\hat Q_{N}^\dagger$ (and thus $\hat Q_N$) defined through its action on states of the path basis it seems natural to ask how it acts on simple spin states, i.e. momentum states built from a spin configuration. Generically, the path states are rather complicated superpositions thereof, and hence we have to find the transformation relating the two bases. In order to work it out, we must address the question of incompleteness of the path basis for chains of odd length, pointed out in section \ref{sec:baxterbethe}. In fact, for odd $N$ we have to \textit{define} the action of $\hat Q_{N-1}^\dagger$ on the missing two states which we denote by $|\Psi_\pm\rangle$.

Let us first state a simple observation: as the $|\Psi_\pm\rangle$ are not in the path basis, they cannot be obtained through the action of $\hat Q_N^\dagger$ on any state in the Hilbert space $\mathcal H_{N+1}$ for the chain with $N+1$ sites. Second, we extend the definition of $\hat Q^\dagger_{N-1}$ in the most natural way: $\hat Q^\dagger_{N-1}|\Psi_\pm\rangle = 0$. Of course, the same reasoning applies to the operator $\hat Q_N$ itself, and thus we have
\begin{equation}
  \hat Q_N|\Psi_\pm\rangle = 0,\quad \text{and} \quad |\Psi_\pm\rangle \neq \hat Q_{N-1}|\phi\rangle \quad \text {for all }|\phi\rangle \in \mathcal H_{N-1}.
  \label{eqn:gsprop}
\end{equation}
Notice that this provides a consistent extension of the nilpotency property $\hat Q_{N+1}\hat Q_N=0$. In a more mathematical parlance, we extend thus the definition of $\hat Q_N$ in such a way that the missing states are closed, but not exact with respect to the operators $\hat Q_N$.

Given this extension, the next steps are to construct the relation
between the path basis and the spin basis, and to understand the
action of the operator $\hat Q_N$ on simple spin states. We have not
found a systematic construction for generic $N$. It seems that this is
related to quite non-trivial identities between Jacobi theta
functions, as suggested by the most simple example $\hat Q_2$. We work
out this special case in appendix \ref{app:q32}. It suggests the
following

\conjecture{For any $N$ the operator $\hat Q_N$ is a linear combination of the supercharges $Q_N$ and $\widetilde Q_N$, defined in section \ref{sec:hamiltonian}, with coefficients depending on the free parameters $s$ and $t$. The latter can be fine-tuned in order to make one of the coefficients vanish. As a generalisation of \eqref{eqn:intertwt} we have the intertwining relation
\begin{equation*}
  \bm{\mathcal T}_N(u)Q_{N-1} + h(u)Q_{N-1}\bm{\mathcal T}_{N-1}(u)=0
\end{equation*}
(and a similar relation for $\widetilde Q_N$)on the momentum sectors with $t_N = (-1)^{N+1}$.
}
\medskip

It is easy to check that this equation is compatible with the commutation relations between the supercharges and the XYZ Hamiltonian \eqref{eqn:crhq}. More importantly however, this conjecture -- if true -- has some interesting consequences for the ground states of the XYZ spin chain along the supersymmetric line. In fact, combining it with \eqref{eqn:gsprop} we conclude that the two missing states $|\Psi_\pm\rangle$ correspond to non-trivial elements in the cohomology of the supercharge $Q_N$ (or $\widetilde Q_N$), discussed previously. Therefore they are perfect candidates for the ground states of the XYZ Hamiltonian for chain with odd $N$. Indeed, we verified this conjecture up to $N=9$ sites by checking that the ground states obtained through exact diagonalisation of the Hamiltonian are indeed orthogonal to all states in the path basis provided that we impose the relation between $\zeta$ and the elliptic nome $q$ given in \eqref{eqn:defzeta}. Thus we are led to

\conjecture{For odd $N$ the subspace of vectors that are orthogonal to the path basis is  two-dimensional. It is spanned by the two zero-energy ground states of the XYZ-Hamiltonian \eqref{eqn:xyzham} which are invariant under translation.}
\medskip

This is perhaps the most surprising outcome of our analysis because it suggests 
that for $\eta=\pi/3$ the ground states at odd $N$ cannot directly be obtained from Baxter's Bethe ansatz. Notice that this observation is different from the widely discussed question of the completeness of the Bethe ansatz (see e.g. \cite{baxter:02}) as here the Bethe ansatz (as it stands) does not apply to the missing states.

These two states are thus eigenstates of the eight-vertex transfer
matrix with eigenvalue $\Lambda_N(u) = \vartheta_1(u,q)^N$ as was
conjectured by Stroganov \cite{stroganov:01,stroganov:01_2}. 
The conjecture was extended to the inhomogeneous eight-vertex model,
defined by allowing on any site $j$ a shift of the spectral parameter
 $u \to u-u_j$. The conjectured eigenvalue is \cite{razumov:10} 
\begin{equation}
  \mathcal T_N(u) = \vartheta_1(u-u_1,q)\vartheta_1(u-u_2,q)\cdots \vartheta_1(u-u_N,q).
  \label{eqn:lambdainhom}
\end{equation}
The simple product structure of this eigenvalue suggests that there is
a local mechanism  leading to its existence. Here we seek to extend
our conjecture 4 to the inhomogeneous setting. The construction of the path basis parallels the homogeneous case, with a slight modification of the local vectors \eqref{eqn:factors}. For site $j$ they become
\begin{align*}
  |\Phi_{\ell,\ell+1}^{(j)}\rangle =
    \vartheta_1(s+(2\ell+1)\eta+u_j,q^2)|+\rangle +
    \vartheta_4(s+(2\ell+1)\eta+u_j,q^2)|-\rangle,\\
    |\Phi_{\ell+1,\ell}^{(j)} \rangle =
    \vartheta_1(t+(2\ell+1)\eta-u_j,q^2)|+\rangle +
    \vartheta_4(t+(2\ell+1)\eta-u_j,q^2)|-\rangle,
\end{align*}
where $|\pm\rangle$ are the local spin$-1/2$ basis vectors.
Using this, we checked numerically for $N=3,5$ and $7$ sites, and random choices for the spectral parameters $u_j$ the following 
\conjecture{For odd $N$ the subspace of vectors that are orthogonal to the inhomogeneous path basis is two-dimensional. It is spanned by the two eigenstates of the inhomogeneous transfer matrix with eigenvalue \eqref{eqn:lambdainhom}.}

\section{Relation to lattice fermions with hard-core exclusion}
\label{sec:fermions}
In this section, we present new observations about the connection
between the XYZ model along the supersymmetric line and the staggered
supersymmetric fermion chains with nearest-neighbour exclusion
considered in \cite{fendley:10,fendley:10_1} (see also
\cite{huijse:11_1}). We provide a first step to construct a mapping
between the models, based on the path description of the states in
Baxter's Bethe ansatz for the eight-vertex model. Such a mapping was
relatively straightforward to obtain for the XXZ case
\cite{fendley:03,yang:04}, because in both cases there is a conserved
$U(1)$ symmetry, and have closely related Bethe equations. Here there
remains a $U(1)$ symmetry in the fermion model, but there is no such
manifest symmetry in the XYZ chain. Nevertheless, there appears to be
a close relation between the spectra in the two cases. Moreover, we
explain how there is evidence of the hard-fermion structure in the
path basis of the eight-vertex model by exploiting the $\text{mod
}3$-periodicity of the heights in the path description.

\subsection{Conjectures relating the spectra}

Let us recall the model of \cite{fendley:03_2,fendley:03}, describing
spinless fermions on a periodic one-dimensional lattice with $N_{(f)}$
sites. The fermions are subject to the constraint that no two adjacent
sites are both occupied. Defining the ordinary fermion creation and
annihilation operators to be $c_j^\dagger$ and $c_j$ with
$\{c_i,c_j\}=\{c_i^\dagger,c_j^\dagger\}=0$ and
$\{c_i,c_j^\dagger\}=\delta_{ij}$, the constraint amounts to
restricting the usual fermionic Hilbert space to states annihilated by $n_jn_{j+1}$,
where $n_j= c_j^\dagger c_j$ is the fermion number
operator. Fermions respecting this constraint are
annihilated and created by the operators $d_j =
(1-n_{j-1})c_j(1-n_{j+1})$ and $d_j^\dagger =
(1-n_{j-1})c_j^\dagger(1-n_{j+1})$.  The model has explicit
$\mathcal N=2$ supersymmetry: it is built from a supercharge
\begin{equation*}
  Q_{(f)}=\sum_{j=1}^{N_{(f)}} \lambda_j d_j^\dagger,
\end{equation*}
where the $\lambda_j$ are non-zero real coupling constants (possible
phases may be removed by simple gauge transformations. The Hamiltonian
is given as anticommutator $H_{(f)} =
\{Q_{(f)},Q_{(f)}^\dagger\}$. For periodic boundary conditions
$d_j=d_{j+N_{(f)}}$ on the fermions (known as \textit{Ramond boundary
conditions}), the Hamiltonian is 
\begin{equation*}
  H_{(f)} = \sum_{j=1}^{N_{(f)}}
  \lambda_j\lambda_{j+1}(d_{j+1}^\dagger d_j + d_j^\dagger
  d_{j+1})+\sum_{j=1}^{N_{(f)}}\lambda_j^2(1-n_{j-1})(1-n_{j+1})\ .
\end{equation*}
This thus includes a hopping term, a chemical potential, and a
next-to-nearest neighbour repulsion. Notice that unlike in the XYZ
chain where the magnetisation is not conserved, this Hamiltonian
conserves fermion number for any values of the $\lambda_j$.

Following \cite{fendley:10,fendley:10_1} we now consider the case where length of the fermion chain is a multiple of three, and the coupling constants are staggered with period three. Then the problem is invariant under translation by three sites: if $T_{(f)}$ is the translation operator on the fermion chain then we have $[H_{(f)},T_{(f)}^3]=0$. In this case, one can show that for $N_{(f)}=3m$ the model has exactly two zero-energy ground states with $m$ fermions in the ``momentum sector'' where $T_{(f)}^3\equiv(-1)^{m+1}$. The precise form of these ground states depends on the values of $\lambda_1,\lambda_2,\lambda_3$. The most general case is analysed in \cite{blom:11}. Here we describe the choice $\lambda_1=y,\lambda_2=1,\lambda_3=y$ for some real $y$. This was the case studied in \cite{fendley:10_1}, where we conjectured that after the change of variable
\begin{equation}
  \zeta^2 = 1+8y^2
  \label{eqn:chgvar}
\end{equation}
the two zero-energy ground states of the fermion chain at $N_{(f)}=3m$, and the two zero-energy ground states of the XYZ spin chain with $N=2m+1$ sites share some components which are polynomials in $\zeta$ and are related to a tau-function hierarchy associated with the Painlev\'e VI equation \cite{bazhanov:05,bazhanov:06,mangazeev:10}.

It is natural to ask if the relation between the two models is deeper. Indeed, if we rewrite the spectrum of the fermion chain in terms of the variable $\zeta$ by using \eqref{eqn:chgvar} then a number of eigenvalues in the spectra of the XYZ Hamiltonian at $N=2m$ and $N=2m+1$ coincide \textit{exactly} with eigenvalues of the fermion chain Hamiltonian $4 H_{(f)}$ (the factor $4$ is just an issue of normalisation). This statement can be sharpened by analysing different momentum sectors. As an example, we provide the characteristic polynomial $\det(E-H_N)$ for the XYZ Hamiltonian with $N=4$ sites in the subsector with momentum $k=\pi$:
\begin{equation}
  \left(E-(\zeta ^2+3)\right) \left(E-(\zeta ^2+2 \zeta +5)\right) \left(E-(\zeta ^2-2 \zeta +5)\right) \left(E-2 \left(\zeta ^2+1\right)\right).
  \label{eqn:cpxyzN4k0}
\end{equation}
The characteristic polynomial $\det(\epsilon -H_{(f)})$ for the fermion model at $N_{(f)}=6$ sites in the subsector with $m=2$ particles and $T_{(f)}^3=-1$ is given by:
\begin{equation}
  \epsilon ^2 \left(\epsilon-(1+4 y^2) \right) \left(\epsilon -(1+2 y^2)\right) \left(\epsilon ^2-\left(3+4 y^2\right) \epsilon+2(1+2 y^2+2 y^4 )\right).
  \label{eqn:cpssfN6m2}
\end{equation}
If we set $\epsilon = E/4$ and use the change of variables \eqref{eqn:chgvar}, then \eqref{eqn:cpssfN6m2} coincides with
\eqref{eqn:cpxyzN4k0} up to the factor $E^2$ and an unimportant global numerical factor. Hence we see that upon the change of variables, the spectra coincide with the exception that the zero-energy states are absent in the XYZ spectrum.
This coincidence of the XYZ spectrum at $N=2m$ on the sector with momentum $\pi$, and the fermion model at $N_{(f)} = 3m$  on the sectors with $T_{(f)}^3=(-1)^{m+1}$ appears to be systematic for small $m$, but different multiplicities of various eigenvalues occur for $m\geq 4$. Studying the spectra up to $m=6$, we are led to the the following conjecture:

\conjecture{The spectrum of the XYZ Hamiltonian $H_N$ for $N=2m$ sites in the sector with momentum $\pi$ coincides with the spectrum of the staggered fermion chain $4H_{(f)}$ with $N_{(f)}=3m$ sites in the sector where $T_{(f)}^3 =(-1)^{m+1}$ if variables are changed according to \eqref{eqn:chgvar}, with two exceptions: \textit{(i)} the eigenvalue $E=0$ is missing in the XYZ spectrum and \textit{(ii)} the two models lead to different multiplicities of the eigenvalues.
\label{conj:relrm}
} \\

This conjecture identifies sectors of the two models where their
supersymmetries are exactly realised. Yet, it appears that the
connection is even deeper. We analysed the relations between the
models for antiperiodic or \textit{Neveu-Schwarz boundary conditions}
$d_{j+N_{(f)}}=-d_j$ on the fermions. In this case, the supersymmetry
of the fermion model is broken. The spectrum needs no longer be
positive, and indeed the ground state has negative energy. These boundary conditions are equivalent to a twist in the Hamiltonian, leading to the term $-\lambda_{N_{(f)}}\lambda_1(d_{1}^\dagger d_{N_{(f)}}+ d_{N_{(f)}}^\dagger d_{1})$. This sector is unlikely to share properties with the momentum sectors discussed so far in this paper because they have explicit unbroken supersymmetry. We found however coincidence with the spectrum of the XYZ chain of even length $N=2m$ and \textit{zero} momentum. We illustrate it once again by showing the explicit characteristic polynomials for $m=2$. The XYZ Hamiltonian for $N=4$ sites, and momentum $k=0$ has the characteristic polynomial
\begin{equation*}
  (E-4) \left(E-(\zeta-1)^2\right) \left(E-(\zeta+1)^2\right) \left(E^3-3 E^2 \left(\zeta ^2+1\right)+2 E \left(\zeta ^2+3\right)^2+8 \left(\zeta ^2-1\right)^2\right)
\end{equation*}
The characteristic polynomial of the fermion model at $N_{(f)}=6$ with Neveu-Schwarz boundary conditions restricted to the sector $T_{(f)}^3=1$ is given by
\begin{equation*}
  (\epsilon-1) \left(\epsilon^3-3 \epsilon^2 \left(2 y^2+1\right)+2 \epsilon \left(2 y^2+1\right)^2+8 y^4\right) \left(\epsilon^2-\epsilon \left(4 y^2+1\right)+4 y^4\right).
\end{equation*}  
Again, if we set $\epsilon = E/4$ and perform the change of variables \eqref{eqn:chgvar} we find that the two polynomials coincide up to some unimportant numerical factor. Studying small systems up to $m=6$ we are led to the 

\conjecture{The spectrum of the XYZ Hamiltonian for $N=2m$ sites in the sector with \textit{zero} momentum coincides with the spectrum of $4H_{(f)}$ for the \textit{twisted} staggered fermion chain with $N_{(f)}=3m$ sites in the sector where $T_{(f)}^3 =(-1)^{m}$ after changing variables according to \eqref{eqn:chgvar}. The multiplicities of the eigenvalues in the two models are different.
\label{conj:relns}
}

\subsection{A mapping to hard-particle configurations}
\label{sec:ssf}
The conjectures \ref{conj:relrm} and \ref{conj:relns}, relating the
spectra of the XYZ chain and the staggered fermion model, raise naturally the question if there is a mapping between the models, at least in some subsectors. At the XXZ point $\zeta=0$ such a mapping was discussed in \cite{fendley:03}: the fermion chain with $N_{(f)}$ sites and $m$ fermions is equivalent to the twisted spin chain with $N=N_{(f)}-m$ sites and $m$ spins down, the twist being the eigenvalue of the translation operator in the fermion model. If we represent an occupied site on the fermion chain by $\bullet$ and an empty site by $\circ$ then the correspondence between fermions and spins is given by
\begin{figure}[h!]
  \centering
  \begin{tikzpicture}
  \begin{scope}
  \clip (0,-0.5) rectangle (1,0.5);
  \draw (0,0) -- (1,0);
  \filldraw[fill=white] (0,0) circle (2pt);
  \filldraw (0.5,0) circle (2pt);
  \filldraw[fill=white] (1,0) circle (2pt);
  \end{scope}
  \draw (2,0) node {$\leftrightarrow$};
  \draw (3,0) node {$-$};
  \draw (4.5,0) node {and};
  
  \begin{scope}
  \clip (6,-0.5) rectangle (6.5,0.5);
  \draw (6,0) -- (6.5,0);
  \filldraw[fill=white] (6,0) circle (2pt);
  \filldraw[fill=white] (6.5,0) circle (2pt);
  \end{scope}
  \draw (7.5,0) node {$\leftrightarrow$};
  \draw (8.5,0) node {$+$};
  \end{tikzpicture}
\end{figure}

This mapping has no direct generalisation to the off-critical case. The reason is the absence of conservation of the number of down spins in the general XYZ chain as opposed to the particle conservation in the staggered fermion chain. However, the path basis was designed to implement particle conservation (the number of down steps). Thus, we focus on the path states, and try to conceive a mapping between them and the fermion model.

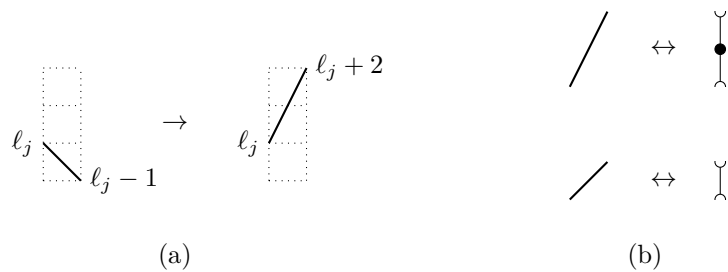
\begin{figure}[h]
  \centering
  \begin{tikzpicture}[>=stealth]
     \draw[scale=0.5,thick] (0,1)--(1,0);
     \draw[scale=0.5,dotted] (0,0) grid (1,3);
     \draw[scale=0.5] (0,1) node[left] {$\ell_j$};
     \draw[scale=0.5] (1,0) node[right] {$\ell_j-1$};
     
     \draw (1.75,0.75) node {$\rightarrow$};

     \draw[xshift=3cm, scale=0.5,thick] (0,1)--(1,3);    
     \draw[xshift=3cm,scale=0.5,dotted] (0,0) grid (1,3);
     \draw[xshift=3cm,scale=0.5] (0,1) node[left] {$\ell_j$};
     \draw[xshift=3cm,scale=0.5] (1,3) node[right] {$\ell_j+2$};
     
     \draw(1.75,-1) node {(a)};
      \draw(8,-1) node {(b)};
    
     \begin{scope} [xshift=7cm,yshift=.75cm]
       \draw[scale=0.5,thick] (0,1)--(1,3);
       \draw (1.25,1) node {$\leftrightarrow$};
       \clip (1.75,0.5) rectangle (2.25,1.5);
       \draw (2,0.5)--(2,1.5);
	   \filldraw (2,1) circle (2pt);   
	   \filldraw[fill=white] (2,1.5) circle (2pt);   
	   \filldraw[fill=white] (2,0.5) circle (2pt);
	 \end{scope}
	 \begin{scope}[xshift=7cm,yshift=.75cm]
	   \draw[scale=0.5,thick] (0,-2)--(1,-1);
       \draw (1.25,-0.75) node {$\leftrightarrow$};
       \clip(1.75,-1) rectangle (2.25,-0.5);
       \draw(2,-1)--(2,-0.5);
	   \filldraw[fill=white] (2,-1) circle (2pt);   
	   \filldraw[fill=white] (2,-0.5) circle (2pt);
	 \end{scope}	  
  \end{tikzpicture}
  \caption{(a) Local modification of the path. (b) Correspondence between path steps and particle configurations.}
  \label{fig:trsfrules}
\end{figure}

\begin{figure}[h]
  \centering
   \begin{tabular}{cc}
  \begin{tikzpicture}[>=stealth]
    \draw (0.3,4.8) node {$\ell_j$};
    \draw[scale=0.5,dotted] (0,-1) grid (7,9);
    \draw[scale=0.5,thick,color=gray] (0,0)--(2,2)--(3,1)--(5,3)--(6,2)--(7,3);
    \draw[scale=0.5,thick] (0,0)--(2,2)--(3,4)--(5,6)--(6,8)--(7,9);
    \draw[scale=1] (4.5,0.0) circle (2pt);
    \draw[scale=1] (4.5,0.5) circle (2pt);
    \draw[scale=1] (4.5,1) circle (2pt);
    \filldraw[scale=1] (4.5,1.5) circle (2pt);
    \draw[scale=1] (4.5,2) circle (2pt);
    \draw[scale=1] (4.5,2.5) circle (2pt);
    \draw[scale=1] (4.5,3) circle (2pt);
    \filldraw[scale=1] (4.5,3.5) circle (2pt);
    \draw[scale=1] (4.5,4) circle (2pt);
    \draw[scale=1] (4.5,4.5) circle (2pt);
    \filldraw[color=white] (4.3,0) rectangle (4.7,-0.2);
    \filldraw[color=white] (3.3,4.5) rectangle (4.7,4.7);
    \draw[<->] (0,5)--(0,0)--(4,0);
    
    \draw(2,-1) node {(a)};
    
    \draw (-0.1,0) node[left] {$\ell$};
     \draw (4,-0.3) node{$j$};
    \draw (0.25,-0.3) node{$1$};
    \draw (3.25,-0.3) node{$N$};
    
       \foreach \x in {0,0.5,...,3}
      \draw[xshift=\x cm] (0.25,0.075)--(0.25,-0.075);
    \foreach \x in {0,0.5,...,4.5}
      \draw[yshift=\x cm] (0.075,0)--(-0.075,0);  
  \end{tikzpicture}
  &
  \begin{tikzpicture}[>=stealth]
      \draw (0.3,4.8) node {$\ell_j$};
    \draw[scale=0.5,dotted] (0,-1) grid (7,9);
    \draw[scale=0.5,thick,color=gray] (0,0)--(2,2)--(5,-1)--(6,0);
    \draw[scale=0.5,thick] (0,0)--(2,2)--(5,8)--(6,9);
    \draw[scale=1] (4.5,0.0) circle (2pt);
    \draw[scale=1] (4.5,0.5) circle (2pt);
    \draw[scale=1] (4.5,1) circle (2pt);
    \filldraw[scale=1] (4.5,1.5) circle (2pt);
    \draw[scale=1] (4.5,2) circle (2pt);
    \filldraw[scale=1] (4.5,2.5) circle (2pt);
    \draw[scale=1] (4.5,3) circle (2pt);
    \filldraw[scale=1] (4.5,3.5) circle (2pt);
    \draw[scale=1] (4.5,4) circle (2pt);
    \draw[scale=1] (4.5,4.5) circle (2pt);
    \filldraw[color=white] (4.3,0) rectangle (4.7,-0.2);
    \filldraw[color=white] (3.3,4.5) rectangle (4.7,4.7);
    \draw[<->] (0,5)--(0,0)--(4,0);
    
        \draw(2,-1) node {(b)};

    \draw (-0.1,0) node[left] {$\ell$};
    \draw (4,-0.3) node{$j$};
    \draw (0.25,-0.3) node{$1$};
    \draw (3.25,-0.3) node{$N$};
    
    \foreach \x in {0,0.5,...,3}
      \draw[xshift=\x cm] (0.25,0.075)--(0.25,-0.075);
    \foreach \x in {0,0.5,...,4.5}
      \draw[yshift=\x cm] (0.075,0)--(-0.075,0);  

  \end{tikzpicture}
  \end{tabular}
  \caption{(a) Mapping from path configuration to a particle configuration with hard-core exclusion. (b) Insertion of a particle through local operation of the supercharges for the XYZ chain.}
  \label{fig:pathvsfermions}
\end{figure}
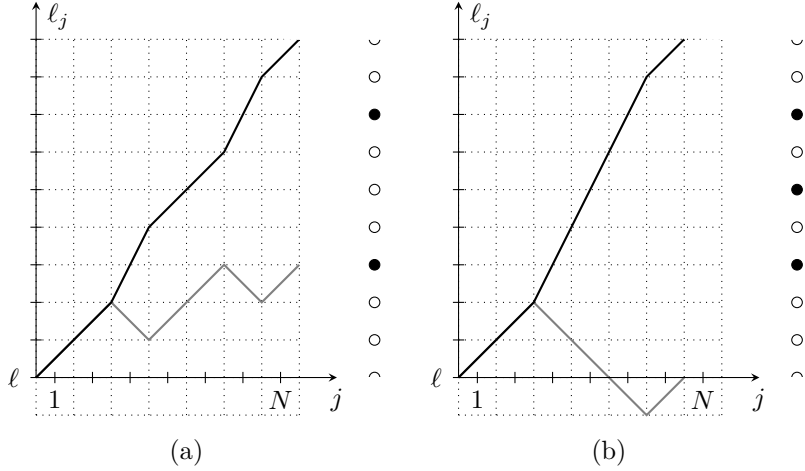

Let us consider a typical path starting at height $\ell_1=\ell$ and
terminating at some $\ell_{N+1}=\ell +3p$ for fixed integer $p$. As
before, let $m$ be the number of decreasing steps. Recall that
adjacent heights obey $\ell_{j+1}-\ell_j = \pm 1$. However, notice
that at $\eta=\pi/3$ we may shift any local height variable by a
multiple of three without changing the corresponding state, as can be
seen from the vectors \eqref{eqn:factors}. Thus, instead of a
decreasing step $\ell_{j+1}=\ell_j-1$ we can modify the path locally
according to $\ell_{j+1} = \ell_j+2$ as shown in figure
\ref{fig:trsfrules}(a). This motivates the following construction:
given a path we replace each decreasing step by a step of two units
up, and then continue with usual, appropriately shifted steps up. This
procedure yields a new, monotone increasing path from height $\ell$ to
height $\ell + N + m +1$ as illustrated in figure
\ref{fig:pathvsfermions}(a). Next, we associate to the each of the two
types of steps particle configurations along the vertical axis
according to the rules display in figure \ref{fig:trsfrules}(b). This
is quite reminiscent of the correspondence in the critical case. Thus,
we obtain from a path a particle configuration with $N_{(f)}=N+m$
sites and $m$ particles (see figure \ref{fig:pathvsfermions}(a) for
illustration), with the hard-core rule that particles cannot be
adjacent to each other, just like in the fermionic case. Notice in
particular that because of the condition $N=2m+3p$, the length of the
particle chain $N_{(f)}=3(m+p)$ is always a multiple of three.

For fixed $\ell$, the position of the particles $\bullet$ on the vertical axis are given as
\begin{equation*}
  y_j = x_j+\ell + j, \quad j=1,\dots,m.
\end{equation*}
We fix the origin at $\ell=y=0$, and consider periodic boundary
conditions, so the $y_j$'s are considered $\text{mod}\, N_{(f)}$.

The mapping between paths and particle configurations is not one-to-one. We illustrate this by analysing the effect of a translation on the path configuration by one step to the right. The last step of the path is simply removed, and glued to the first one. However, we would like to respect the rule that the initial height of the new path is $0,1$ or $2$. Thus, a vertical shift of all heights by $\pm 3$ units might be necessary, and lead to a different particle configuration. There are multiple cases. \textit{(i)} If the last step of the given path goes up, and $\ell=1,2$ the translation has no effect on the configuration of hard-particles.  \textit{(ii)} However, if $\ell=0$ the path has to be shifted by three units, and thus the particle positions are cyclically translated by $3$: $y_j\to y_j+3 \mod N_{(f)}$. \textit{(iii)} If the last step of the initial path is decreasing, then a translation of the path leaves the particle configuration unchanged if $\ell=2$. \textit{(iv)} For $\ell = 0,1$ however, the positions in the particle configuration constructed from the translated path are shifted according to $y_j\to y_j+3 \mod N_{(f)}$. For given $N$ and paths with $m$ steps down, the number of hard-particle configurations obtained through the mapping is obtained by counting the paths corresponding to cases \textit{(ii)} and \textit{(iv)}. This yields
\begin{equation*}
  {N-1 \choose m} + 2 {N-1 \choose m-1}= \frac{N_{(f)}}{N_{(f)}-m}{N_{(f)}-m \choose m},
\end{equation*}
where we used $N_{(f)}=N+m$. A little combinatorics shows that this is the number of possible hard-particle configurations for $N_{(f)}$ sites and $m$ particles. Also, we see that translation of the path configurations is related to translation of the hard-particle configurations along the vertical axis by three steps.

Next, we would like to understand the nature of these particles by
examining the local action of the XYZ supercharges, explained in
section \ref{sec:derivsusy}, on the particle configurations. Recall
that the action on the path states corresponds locally to transforming
two up steps into a single down step. Given our rules identifying
paths with particle configurations, it is not difficult to see that
this corresponds to insertion of a particle while respecting the
nearest-neighbour exclusion rule (see figure
\ref{fig:pathvsfermions}(b)). Recall that this comes with a weight
$(-1)^x h(w_{\ell_{x+1}})^2$ in terms of the positions for the
path. The corresponding position of insertion in the hard-particle
state is $y=x+\ell +j$ where $j$ is the number of particles on sites
$0,1,\dots,y-2$. Hence, the weight becomes $(-1)^{j+y-\ell} h(w_y)^2$
as follows from \eqref{eqn:defw} and the periodicity of $h(u)$. The
factor $(-1)^j$ in the string suggests that the hard particles are
indeed fermions. Furthermore it is tempting to use this in order to
identify the coupling constants $\lambda_y$. This requires taking into
account a systematic identification of the hard-particle states in
terms of the path states (what is delicate as the proposed mapping is
not one-to-one), changes of normalisations through the supersymmetry
operation, and finally the restriction to the special momentum spaces
for both models. While we are not in a position to carry out this program, we nevertheless put forward the following
\conjecture{The coupling constants of the corresponding fermion model are given by
\begin{equation*}
  \lambda_y = |\vartheta_1(w_{y},q)|^{3/2}.
\end{equation*}
where $w_y$ is the linear function defined in \eqref{eqn:defw}.}\\

We see that these coupling constants depend only on the combination $s+t$. The evidence for this conjecture is that this parametrisation of the coupling constants uniformises a family of elliptic curves appearing in the coordinate direct coordinate Bethe ansatz for the fermion chain \cite{blom:11}. In particular, it implies that upon appropriate rescaling the eigenvalues of the fermion chain do not depend on $s+t$.

\section{Conclusion}
\label{sec:concl}
We have studied the XYZ chain and the eight-vertex model along the
supersymmetric line, and showed that it possesses a $\mathcal N=(2,2)$
supersymmetry on the lattice. A consequence is that chains of
different length have common positive energy levels in certain
momentum sectors, which are organised into supersymmetry quadruplets. Moreover, we presented a derivation of the supersymmetry by means of the Bethe ansatz for the eight-vertex model, and showed that the supercharges perform simple local operations on the path basis. This analysis led us to a novel characterisation for the ground states of the XYZ chain with odd length. Finally, we reported some observations that the XYZ chain along the supersymmetric line and the staggered supersymmetric fermion chains with nearest-neighbour exclusion have exact common eigenvalues in certain subsectors.

There are many open questions and extensions. To us, it seems most
interesting to clarify further the nature of the ground states for the
chains of odd length. We hope that the supersymmetry will be helpful,
for instance to prove that there are exactly two zero-energy ground
states. A central tool in supersymmetric theories is the Witten index
$\tr (-1)^F$ \cite{witten:82}: it provides a lower bound on the
number of zero-energy states. Indeed, it would be interesting to
define this quantity or find at least a suitable analogue for the
present theory. As the fermion number coincides with the number of
sites, the formal generalisation leads to a trace which runs over an
infinite collection of Hilbert spaces, what one would have to make
sense of. Similar considerations apply to the index $\tr ((-1)^F F
e^{-\beta H})$
defined in \cite{cecotti:92}. A possible way to resolve these problems
might be to establish a more complete mapping between the XYZ chain and the
staggered fermion chain. For the latter, there exists a standard
procedure to find the Witten index, and determine the exact number of
ground states using cohomology arguments (see e.g. \cite{huijse:10_1}). Further
insights into the structural properties of the ground states will
certainly be obtained by considering the inhomogeneous eight-vertex
model, as was the case in the trigonometric limit
\cite{difrancesco:04,difrancesco:04_2,difrancesco:05_3,difrancesco:06}. Almost
all developments in this work considered the homogeneous version, and
the supersymmetry appears to be intimately related to translation
invariance. It would be interesting to see if (and how) this symmetry
persists in the inhomogeneous case.

Finally, let us point out that the supersymmetry presented in this
article is a particular feature of the $\eta=\pi/3$ model. It is
natural to ask for an extension to general roots-of-unity points such
as $\eta=\pi/(k+2)$ with $k=1,2,3,\dots$ The case $k=2$ was already
addressed in \cite{fendley:03} from the point of view of fermions with
generalised exclusion rules. This allowed the identification of a
supersymmetric point for the Fateev-Zamolodchikov integrable spin-$1$
chain. For more general trigonometric models, the points
$\eta=\pi/(k+2)$ were identified as the combinatorial points for fused
spin-$k/2$ models \cite{zinn:09}, as anticipated in \cite{dorey:04} (see also \cite{saleur:92}). Indeed, using the works \cite{kirillov:87,takebe:95} we can show that these coincide precisely with the cases where a lattice supersymmetry is present. This generalisation will be addressed in a forthcoming publication \cite{hagendorf:2bp}.

\subsection*{Acknowledgements} This work was supported by the NSF
grants DMR/MPS-0704666 and 1006549. CH was supported in part by the ERC AG CONFRA and by the Swiss NSF. He acknowledges support from the Kavli Institute for Theoretical Physics under NSF grant PHY05-51164, and would like to thank the KITP, where most of this work was done, for hospitality. Moreover, CH would like to thank Luc Blom, Gaetan Borot, Bernard Nienhuis, and Paul Zinn-Justin  for very fruitful discussions, and Liza Huijse and Anton Zabrodin for discussions at early stages of this work.

\appendix

\section{Properties of the supercharges}
In this appendix we present some technical details about the properties of the supercharges introduced in section \ref{sec:hamiltonian}. In the first part, we prove the nilpotency property, and in the second part, we show that their anticommutator generates the XYZ Hamiltonian.

\subsection{Nilpotency}
\label{app:nilpotency}
Let us prove that the operators $Q_N$ ``have square zero'' in the sense that
\begin{equation*}
  Q_{N+1}Q_N=0.
\end{equation*}
To this end, we need a set of anticommutation rules for the local operators $q_j$ defined in the main text \eqref{eqn:defq}.
We have the rule
\begin{equation}
  q_i q_j + q_{j+1} q_i=0, \quad 1\leq i<j\leq N. \label{eqn:acqj}
\end{equation}
This can be shown along the lines of \cite{yang:04}, and therefore we only sketch the proof for another relation involving $q_0$. Let us first consider $q_0 q_j$. We find that its action non-zero only on states having spins $-$ at position $j$ and $N$. We find
\begin{align*}
  q_0 q_j |\cdots \underset{j}{-} \cdots \underset{N}{-}\rangle =(-1)^j&\bigl(|+\cdots \underset{j+1}{+} + \cdots \underset{N+2}+\rangle-\zeta|+\cdots \underset{j+1}{-} - \cdots \underset{N+2}+\rangle\\
  &-\zeta|-\cdots \underset{j+1}{+} {+} \cdots \underset{N}-\rangle+\zeta^2|-\cdots \underset{j+1}{-} - \cdots \underset{N}-\rangle\bigr).
\end{align*}
Reversing the order of the $q$'s, we have to take into account the shift and therefore consider $q_{j+1} q_0$. Its action yields
\begin{align*}
   q_{j+1} q_0 |\cdots \underset{j}{-} \cdots \underset{N}{-}\rangle =(-1)^{j+1} &\bigl(|+\cdots \underset{j+1}{+} + \cdots \underset{N}+\rangle-\zeta|+\cdots \underset{j+1}{-} -\cdots \underset{N}+\rangle\\
  &-\zeta|1\cdots \underset{j+1}{+} + \cdots \underset{N}-\rangle+\zeta^2|-\cdots \underset{j+1}{-} - \cdots \underset{N}-\rangle\bigr).
\end{align*}
We see that this result coincides with the previous one, except for a minus sign.
Combining these two equations, we find therefore
\begin{equation}
  q_0 q_j+q_{j+1}q_0 =0, \quad j=1,\dots,N-1 \label{eqn:acqhat}
\end{equation}
when acting on $\mathcal H_N$.

These relations are useful in order to prove that the supercharges are nilpotent in the sense stated above.
In a first step, we observe that the \eqref{eqn:acqj} and \eqref{eqn:acqhat} can be used to reduce the product of the supercharges to
\begin{equation*}
  Q_{N+1}Q_{N}=\left(\frac{N}{N+2}\right)^{1/2}\left(\sum_{j=0}^{N} \left(q_{j+1}q_j + q_j^2\right)+ q_0 q_{N} + q_{N+1} q_0 \right).
\end{equation*}
Let us examine the different terms in this sum. The individual terms are non-vanishing only if they act on the following states in $\mathcal H_{N}$:
\begin{align}
  \left(q_{j+1}q_j + q_j^2\right)|\cdots \underset{j}{-}\cdots\rangle &=\zeta\left(|\cdots \underset{j}{-}+\hspace{-0.45em}\underset{j+2}{+}\cdots \rangle-|\cdots \underset{j}{+}+\hspace{-0.45em}\underset{j+2}{-}\cdots \rangle\right),\quad j=1,\dots,N\nonumber \\
  \label{eqn:defrules}
  \left(q_1 q_0+q_0^2\right)|\cdots \underset{N}{-}\rangle &= \zeta\left(|\underset{1}{-}-\cdots \underset{N+2}{-}\rangle-|\underset{1}{+}-\cdots\underset{N+2}+ \rangle\right),\\
  \left(q_0 q_{N} + q_{N+1} q_0\right)|\cdots \underset{N}{-}\rangle &= (-1)^{N+1}\zeta\left(|\underset{1}{+}\cdots \underset{N+1}{-}+ \rangle-|\underset{1}{-}\cdots \underset{N+1}{+}\hspace{-0.5em}+ \rangle\right).\nonumber
\end{align}
As all the expressions are proportional to $\zeta$ we see that in the XXZ limit $\zeta=0$ the relation $Q_{N+1}Q_N=0$ is immediate. Actually, it would not even be necessary to impose the restriction to certain momentum spaces in this case. However, for general $\zeta\neq 0$ the relation only survives on the special momentum sectors. Intuitively, this can be seen as follows: we see that the operations defined in \eqref{eqn:defrules} insert pairs $++$ to left and right of a spin $-$, thus we expect that the summation of these on a periodic chain will lead to telescopic cancellations. This will however only work in a momentum sector compatible with the sign in the third expression in \eqref{eqn:defrules} which coincides with the eigenvalue of the translation operator. More concretely, let us consider the case of $Q_{N+1}Q_{N}$ on a momentum state $|\psi_\alpha\rangle$ built from spin configuration $\alpha=\alpha_1\alpha_2\cdots\alpha_N$, that is
\begin{equation*}
  |\psi_\alpha\rangle = \sum_{j=0}^{N}t^j_N T_N^{-j}|\alpha\rangle, \quad t_N=(-1)^{N+1}.
\end{equation*}
We leave aside the issue of normalisation. Let us suppose that $\alpha$ has $m$ spins $-$ and denote their positions by $x_1,\dots, x_m$. Using the rules defined in \eqref{eqn:defrules} we can write
\begin{align*}
  C\zeta^{-1}Q_{N+1}Q_{N}|\psi_\alpha\rangle = &\sum_{j=0}^{N-1}t_N^j(T_N^{-j}|\alpha\rangle\otimes |{+}{+}\rangle-|{+}{+}\rangle\otimes T_N^{-j}|\alpha\rangle)\\
  & +(-1)^{N+1}\sum_{\ell=1}^m t_N^{x_\ell}\left(|{+}\rangle\otimes T_N^{-x_\ell}|\alpha\rangle\otimes |{+}\rangle- T_N^{1-x_\ell}|\alpha\rangle\otimes |{+}{+}\rangle\right)\\
  &+\sum_{\ell=1}^m t_N^{x_\ell}\left(|{+}{+}\rangle\otimes T_N^{-x_\ell}|\alpha\rangle- |{+}\rangle \otimes T_N^{1-x_\ell}|\alpha\rangle\otimes |{+}\rangle\right).
\end{align*}
where $C=\sqrt{(N+2)/N}$.
Because of $t_N=(-1)^{N+1}$ this can be written in terms of the positions $y_1,\dots,y_{N-m}$ of the individual spins $+$ in $\alpha$. We find a simplified expression:\begin{align}
 C\zeta^{-1}Q_{N+1}Q_{N}|\psi_\alpha\rangle = & (-1)^{N}\sum_{\ell=1}^{N-m} t_N^{y_\ell}\left(|+\rangle\otimes T_N^{-y_\ell}|\alpha\rangle\otimes |+\rangle- T_N^{1-y_\ell}|\alpha\rangle\otimes |{+}{+}\rangle\right)\nonumber\\
  &-\sum_{\ell=1}^{N-m} \nonumber t_N^{y_\ell}\left(|{+}{+}\rangle\otimes T_N^{-y_\ell}|\alpha\rangle- |{+}\rangle \otimes T_N^{1-y_\ell}|\alpha\rangle\otimes |{+}\rangle\right).
\end{align}
Now observe that if the configuration $\alpha$ has a spin $+$ at position $y_i$ we can write
\begin{equation*}
  |+\rangle\otimes T_N^{-y_i}|\alpha\rangle= T_N^{1-y_i}|\alpha\rangle \otimes |+\rangle.
\end{equation*}
Using this in the preceding formula we conclude that all terms cancel mutually. This proves the statement $Q_{N+1}Q_{N}=0$.

\subsection{The Hamiltonian as an anticommutator}
\label{app:hamiltonian}
In this appendix we show in detail that if we restrict the Hamiltonian to subsectors where the eigenvalue of the translation operator is $t_N=(-1)^{N+1}$ then it can be written as ``anticommutator''
\begin{equation}
  H_N = Q_N^\dagger Q_N + Q_{N-1}Q_{N-1}^\dagger.
\end{equation}
First, it is useful to introduce the projector on the momentum spaces that we are interested in. It is given by
\begin{equation*}
  \Pi_N = \frac{1}{N}\sum_{j=0}^{N-1} t_N^j T_N^{j}=\frac{1}{N}\sum_{j=0}^{N-1} (-1)^{(N+1)j} T_N^{j}.
\end{equation*}
From the definition of the supercharges \eqref{eqn:defq} and the translation properties established in \eqref{eqn:trsl}, we conclude that $Q_N\Pi_N =\Pi_{N+1} Q_N = Q_N$. Using this, we conclude that
\begin{align*}
  Q_N^\dagger Q_N =N(N+1) \Pi_N q_1^\dagger\Pi_{N+1}q_1\Pi_N ,\\
  Q_{N-1}Q_{N-1}^\dagger =N(N-1)\Pi_N q_1\Pi_{N-1}q_1^\dagger\Pi_N .
\end{align*}
Second, using the definition of the projector, and again the rules \eqref{eqn:trsl}, we find that
\begin{align*}
  &(N+1)\Pi_N q_1^\dagger\Pi_{N+1}q_1\Pi_N 
  = \Pi_N \left(\sum_{j=0}^N q_j^\dagger q_1\right)\Pi_N ,\\
  &(N-1)\Pi_N q_1\Pi_{N-1}q_1^\dagger\Pi_{N} 
  = \Pi_N \left(\sum_{j=1}^{N-1}q_1 q_j^\dagger\right)\Pi_N .
\end{align*}
We reduce in a third step the sum of these expressions through an application of the following anticommutation relations
\begin{align*}
& q_i q_j^\dagger +q^\dagger_{j+1} q_i=0, \quad 1\leq i< j-1\leq N-1,\\
& q_0 q_j^\dagger +q^\dagger_{j+1} q_0=0, \quad 2\leq i \leq N-1.  
\end{align*}
These can be derived in a similar way as the relations \eqref{eqn:acqj} and \eqref{eqn:acqhat}.
After some algebra, we are left with
\begin{equation*}
  \Pi_N H_N\Pi_N^{-1} = N\Pi_N \left(q_1^\dagger q_1 +q_1 q_1^\dagger +q_2^\dagger q_1+q_0^\dagger q_1\right)\Pi_N^{-1}.
\end{equation*}
The remaining quadratic terms can be expressed through simple spin operators. We find
\begin{align*}
  q_j q_j^\dagger &= \frac{1}{4}\left((1+\sigma^z_j)(1+\sigma^z_{j+1})+\zeta^2(1-\sigma^z_j)(1-\sigma^z_{j+1})\right) -\zeta\left(\sigma_j^+\sigma_{j+1}^++\sigma_j^-\sigma_{j+1}^-\right),\\
  q_j^\dagger q_j &= \frac{1}{2}(1+\zeta^2)(1-\sigma^z_j),\\
  q_{j+1}^\dagger q_j &=-\frac{\zeta^2}{4}(1-\sigma^z_j)(1-\sigma^z_{j+1})-\sigma_j^+\sigma^-_{j+1},
\end{align*}
and $q_{0}^\dagger q_1 = T_N ^{-1}q_1^\dagger q_2 T_N$.
Using these relations and again translation invariance, we conclude that
\begin{align*}
  \Pi_N H_N\Pi_N^{-1} =&-N\Pi_N \left(\sigma_1^+\sigma_2^-+\sigma_1^-\sigma_2^+ +\zeta(\sigma_1^+\sigma_2^++\sigma_1^-\sigma_2^-)\right)\Pi_N \\
  & -N\Pi_N \left(\left(\frac{\zeta^2-1}{4}\right)\sigma_1^z\sigma_2^z-\frac{3+\zeta^2}{4}\right)\Pi_N .
\end{align*}
The expression on the right-hand side is nothing but the restriction of the XYZ-Hamiltonian (\ref{eqn:xyzham},\ref{eqn:param}) to the momentum sectors with $t_N=(-1)^{N+1}$, what proves the statement.

\section{Reduction from $N=3$ to $N=2$ sites}
\label{app:q32}
In this appendix we show that the supercharge in the path basis can be written as linear superposition of the supercharges defined in section \ref{sec:defsusy} for the most simple case of three and two sites.

There is a single $\pi$-momentum state for $N=2$ sites. In the canonical basis it is given by (we do not normalise the states):
\begin{equation*}
  |\phi\rangle=|{+}{-}\rangle - |{-}{+}\rangle.
\end{equation*}
For $N=3$ sites, there are four possible states which are invariant under translation
\begin{equation*}
  \begin{array}{ll}
  |\psi_1\rangle =|{+}{+}{+}\rangle, &
  |\psi_2\rangle =|{-}{+}{+}\rangle+|{+}{-}{+}\rangle+|{+}{+}{-}\rangle,\\
  |\psi_3\rangle =|{+}{-}{-}\rangle+|{-}{+}{-}\rangle+|{-}{-}{+}\rangle,&  |\psi_4\rangle =|{-}{-}{-}\rangle.
  \end{array}
\end{equation*}

Let us now turn to the path basis. For even $N$ it is redundant. Indeed, for $N=2$ there are $\nu(2)=6$ admissible paths but the Hilbert space has dimension $d=2^N=4$. Indeed, one verifies that the different states are related through the identity
\begin{equation*}
  h(w_{\ell+1})\left(
  \tikz[baseline=1ex]{
    \draw[dotted,scale=0.5] (0,-1) grid (2,1);
    \draw[thick,scale=0.5] (0,0)--(1,1)--(2,0);
    \draw (0,0) node[below] {\small $\ell$};
    \draw (1,0) node[below] {\small $\ell$};
     \draw (0.5,0.5) node[above] {\small $\ell+1$};
  }
  -
  \tikz[baseline=1ex]{
    \draw[dotted,scale=0.5] (0,-1) grid (2,1);
    \draw[thick,scale=0.5] (0,1)--(1,0)--(2,1);
    \draw (0,0.5) node[above] {\small $\ell+1$};
    \draw (1,0.5) node[above] {\small $\ell+1$};
     \draw (0.5,0) node[below] {\small $\ell$};
  }
  \right)
  =
  h(w_{\ell-1})\left(
  \tikz[baseline=-1.1ex]{
    \draw[dotted,scale=0.5] (0,-1) grid (2,1);
    \draw[thick,scale=0.5] (0,-1)--(1,0)--(2,-1);
    \draw (0,-0.5) node[below] {\small $\ell-1$};
    \draw (1,-0.5) node[below] {\small $\ell-1$};
     \draw (0.5,0) node[above] {\small $\ell$};
  }
  -
  \tikz[baseline=-1.1ex]{
    \draw[dotted,scale=0.5] (0,-1) grid (2,1);
    \draw[thick,scale=0.5] (0,0)--(1,-1)--(2,0);
    \draw (0,0) node[above] {\small $\ell$};
    \draw (1,0) node[above] {\small $\ell$};
     \draw (0.5,-0.5) node[below] {\small $\ell-1$};
  }
  \right).
\end{equation*}
In some sense, the relation is trivial here because the difference of the path states on both sides is proportional to the singlet state $|\phi\rangle$, and equality of the proportionality factors on both sides is readily verified.

For $N=3$ there are two states in the path basis which are invariant under translation. Up to factors, they are given by
\begin{equation*}
  |\chi_1\rangle=\sum_{\ell=0}^2
  \tikz[baseline=0.65cm,scale=0.5]
  {
    \draw[dotted] (0,0) grid (3,3);
    \draw[thick] (0,0)--(3,3);
    \draw (0,0) node[below] {\small $\ell$};
    \draw (3,3) node[above] {\small $\ell+3$};
  },\quad |\chi_2\rangle=\sum_{\ell=0}^2
  \tikz[baseline=0.65cm,scale=0.5]
  {
    \draw[dotted] (0,0) grid (3,3);
    \draw[thick] (0,3)--(3,0);
    \draw (0,3) node[above] {\small $\ell$};
    \draw (3,0) node[below] {\small $\ell-3$};
  }.
\end{equation*}
As mentioned before, the path basis for odd $N$ is incomplete. In this concrete example, we see that the two states that are missing have to be invariant under translation. The requirement that they are orthogonal to all path states determines them up to linear combinations and normalisations. We find it convenient to choose
\begin{equation}
  |\chi_3\rangle = \zeta|\psi_1\rangle+|\psi_3\rangle, \quad
  |\chi_4\rangle = \zeta|\psi_4\rangle+|\psi_2\rangle,
  \label{eqn:gsn3}
\end{equation}
where we used the coordinate basis.

In order to find the action on $\hat  Q_2^\dagger$ on the vectors $|\psi_j\rangle$ we write simply decompose the path basis and the two missing states for three sites in terms of the spin basis according to $|\chi_i\rangle=\sum_{j=1}^4 A_{ij}|\psi_j\rangle$. Hence
\begin{equation*}
   \hat Q_2^\dagger |\chi_i\rangle =\sum_{j=1}^4 A_{ij} \hat Q_2^\dagger|\psi_j\rangle = b_i|\phi\rangle,
\end{equation*}
where the $b_i$ are constants (in this example the map is necessarily of rank $1$). Hence
\begin{equation}
   \hat Q_2^\dagger|\psi_i\rangle =\sum_{j=1}^4( A^{-1})_{ij}b_j |\phi\rangle.
   \label{eqn:qcb}
\end{equation}
Thus we have to determine the matrix $A$ involved in the change of basis and the vector $b$. For the former it is convenient to abbreviate $f_j(x) = \prod_{k=0}^2\vartheta_j(x+2\pi k/3,q^2)$. Then we find that
\begin{equation*}
  A=\left(
    \begin{array}{cccc}
      3f_1(s) & -\zeta f_4(s) & -\zeta f_1(s) & 3f_4(s)\\
      3f_1(t) & -\zeta f_4(t) & -\zeta f_1(t) & 3f_4(t)\\
      \zeta & 0 & 1 & 0\\
      0 & 1 & 0 & \zeta    
    \end{array}
  \right),
\end{equation*}
where we used the theta function identity
\begin{align*}
  \vartheta_1\left(u,q^2\right)\left(\vartheta_4\left(u-\frac{\pi}{3},q^2\right)\vartheta_4\left(u+\frac{\pi}{3},q^2\right)+\zeta\vartheta_1\left(u-\frac{\pi}{3},q^2\right)\vartheta_1\left(u+\frac{\pi}{3},q^2\right)\right)\\
 = \vartheta_4\left(u,q^2\right)\left(\vartheta_4\left(u-\frac{\pi}{3},q^2\right)\vartheta_1\left(u+\frac{\pi}{3},q^2\right)+\vartheta_1\left(u-\frac{\pi}{3},q^2\right)\vartheta_4\left(u+\frac{\pi}{3},q^2\right)\right).
\end{align*}
(notice that using the definition of $\zeta$ this turns out to be an identity involving products of \textit{five} theta functions and therefore does not simply follow from Riemann's identity).

Next, let us determine the $b_j$.
The only path state which is not annihilated by $\hat Q_2^\dagger$ is $|\chi_1\rangle$. The application of the local transformation rules defined in \eqref{eqn:defqhat} and \eqref{eqn:weightqhat} leads to
\begin{equation*}
 \hat Q_2^\dagger
 \left(
 \tikz[baseline=0.65cm,scale=0.5]
  {
    \draw[dotted] (0,0) grid (3,3);
    \draw[thick] (0,0)--(3,3);
    \draw (0,0) node[below] {\small $\ell$};
    \draw (3,3) node[above] {\small $\ell+3$};
  }
  \right)
  =-{h(w_{\ell+1})^2} \tikz[baseline]{
    \draw[dotted,scale=0.5] (0,-1) grid (2,1);
    \draw[thick,scale=0.5] (0,0)--(1,-1)--(2,0);
    \draw (0,0) node[above] {\small $\ell$};
    \draw (1,0) node[above] {\small $\ell$};
     \draw (0.5,-.5) node[below] {\small $\ell-1$};
  }
  +{h(w_{\ell-1})^2}  \tikz[baseline]{
    \draw[dotted,scale=0.5] (0,-1) grid (2,1);
    \draw[thick,scale=0.5] (0,0)--(1,1)--(2,0);
    \draw (0,0) node[below] {\small $\ell$};
    \draw (1,0) node[below] {\small $\ell$};
     \draw (0.5,0.5) node[above] {\small $\ell+1$};
  }\end{equation*}
Finally the summation over $\ell =0,1,2$ then yields $b_1= h((s-t)/2)\sum_{\ell=0}^2 h(w_\ell)^3$. From the rules for the action of $\hat Q_2^\dagger$ it is evident that $b_2=0$. Not evident however are the values of $b_3$ and $b_4$. We follow the proposal made in the main text: the supercharges annihilate the two missing states at odd length. Hence we set $b_3=b_4=0$. Then it is a simple matter to find the action of $\hat Q_2^\dagger$ on the spin states. After having computed the inverse $A^{-1}$ of the coordinate transformation we find from \eqref{eqn:qcb}
\begin{equation*}
  \hat Q_2^\dagger = \text{const.}\times\left( f_1(t) Q_2^\dagger + f_4(t)\widetilde Q_2^\dagger\right),
\end{equation*}
with the functions $f_j(t)$ defined above. The overall factor is a function of $s,t$ and $q$. The action of the operators $ Q_2^\dagger$ and $\widetilde Q_2^\dagger$ on the spin states is
\begin{align*}
   Q_2^\dagger|\psi_1\rangle = 0, \quad  Q_2^\dagger|\psi_2\rangle=-\zeta|\phi\rangle, \quad  Q_2^\dagger|\psi_3\rangle = 0,\quad  Q_2^\dagger|\psi_4\rangle=|\phi\rangle,\\
  \widetilde Q_2^\dagger|\psi_1\rangle = -|\phi\rangle, \quad \widetilde Q_2^\dagger|\psi_2\rangle=0, \quad \widetilde Q_2^\dagger|\psi_3\rangle = \zeta|\phi\rangle,\quad \widetilde Q_2^\dagger|\psi_4\rangle=0.
\end{align*}
Thus, we see that $\widetilde Q_2^\dagger = R_2 Q_2^\dagger R_3$ where $R_N$ is the spin-reversal operator introduced in section \ref{sec:notations}. Find with thus the Hermitian conjugates of the supercharges constructed in section \ref{sec:hamiltonian}.

\end{document}